\documentclass[10pt,aps,pra,twocolumn,showpacs,superscriptaddress]{revtex4-1}
\usepackage{amsmath} 
\usepackage{amsthm} 
\usepackage{amssymb} 
\usepackage{color}
\usepackage[bookmarks=false]{hyperref}
\usepackage{graphicx}
\usepackage{hyperref}
\usepackage{dcolumn}
\usepackage{bm}
\usepackage{times}

\newcommand{\bra}[1]{\langle{#1}|}
\newcommand{\ket}[1]{|{#1}\rangle}
\newcommand{\braket}[2]{\langle#1|#2\rangle}
\newcommand{\ketbra}[2]{|#1\rangle_{23}\langle#2|}
\begin{document}
\title{Transmission losses in optical qubits for controlled teleportation}
\author{I. Medina}
\author{F. L. Semi\~ao}
\affiliation{Centro de Ci\^encias Naturais e Humanas, Universidade Federal do ABC, 
             09210-170, Santo Andr\'e, S\~ao Paulo, Brazil}
\begin{abstract}
\noindent In this work, we investigate the controlled teleportation protocol using optical qubits within the single-rail logic. The protocol makes use of an entangled tripartite state shared by the controller and two further parties (users) who will perform standard teleportation. The goal of the protocol is to guarantee that the teleportation is successful only with the permission of the controller. Optical qubits based on either superpositions of vacuum and single-photon states or superposition of coherent states are employed here to encode a tripartite maximal slice state upon which the protocol is based. We compare the performances of these two encodings under losses which are present when the qubits are guided through an optical fiber to the users. Finally, we investigate the non-locality of the shared tripartite state to see whether or not it impacts the efficiency of the protocol.

\end{abstract}
\maketitle
\section{Introduction} 

Entanglement is a property of multipartite quantum systems which possess no complete classical analog. Its existence has been linked to fundamental aspects such as non-locality  \cite{epr,bell} or steering \cite{stee} as well as practical protocols in quantum information \cite{tp,scod}. Quantum teleportation \cite{tp} is a protocol where two parties, Alice (sender) and Bob (receiver), share a bipartite maximally entangled state (the quantum channel) which allows Alice to perfectly send the unknown state of a qubit to Bob by making only local measurements in her laboratory and sending two bits of classical information to him. If the channel is not perfect, i.e., not a pure maximally entangled bipartite state, this protocol allows Alice to just imperfectly or probabilistic send the qubit state to Bob \cite{horodecks2,noisechannel}.

With time, more sophisticated versions of the standard quantum teleportation protocol have been proposed. One example is the so-called controlled teleportation  \cite{ctfirst} (CT) where a third party comes into the play. Charlie (controller) can locally control the quality of the quantum channel shared by Alice and Bob thus affecting the quality of the standard teleportation they intend to perform. To achieve this control, Charlie can share with Alice and Bob a maximally entangled three-qubit Greenberger-Horne-Zeilinger (GHZ) state \cite{GHZ}, and by making or making not a local projective measurement on his qubit, Charlie can influence the fidelity of teleportation \cite{ctfirst}. When Charlie does not perform the measurement, Alice and Bob can only achieve teleportation fidelities up to $2/3$. In practical terms, this means no quantum teleportation since such fidelities can be achieved without the use of any quantum channel \cite{popescu}. In other words, Alice and Bob need the permission of Charlie (his performing a local measurement and announcing the outcome) in order for them to achieve fidelities higher than the classical bound $2/3$. In \cite{gao1}, this problem is studied for a family of pure three-qubit states. In \cite{ghose1}, the so called maximal slice states (MS) are shown to be particularly useful for CT. What is surprising here is that such states are not, in general, maximally entangled tripartite states such as the GHZ state. Both the standard and controlled teleportation protocols find applications in the field of secure quantum communication \cite{security2}.

In the context of quantum optics, different photonic qubits have been considered for practical implementation quantum information protocols. In this work, we will be considering two of such encodings and studying their performance in CT, including realistic losses during the transmission of the qubits.  In the first case, qubit codification is  based on single-mode orthogonal Fock states. The paradigmatic example being the superposition of single-photon and vacuum states. This is known as single-rail quantum logic. Very important developments have been made for such qubits. For instance, a universal set of nondeterministic gates can be built using linear optics and photon counting \cite{lund}.  Improved procedures for performing gates with a significant reduction in the resources have also been discussed \cite{resour}. Next, the use of real-time feedback control was shown to further reduce the resources needed for scalable quantum computing with single-rail optical qubits \cite{feed}. A critical assessment of single-rail quantum linear optical quantum computing can be found in \cite{crit}. Teleportation and violation of local realism for a bipartite extension of this single-rail logic was considered in \cite{pioneer}. From the experimental side, these states have been generated using spontaneous parametric down-conversion \cite{vspimp2} or mixing single-photon and coherent states in a beam splitter, followed by a conditioned measurement \cite{vspimp3}. The teleportation protocol as proposed in \cite{pioneer} has already been experimentally realized \cite{vspimp1}.

In the second case, the codification rely on the use of non-orthogonal single-mode coherent states of opposite phases. It is a generalized type of single-rail qubit in the sense that it tends to the standard case for large coherent state amplitudes. In this limit, the two opposite-phase coherent states used in the encoding approach orthogonality. On the other hand,  for small amplitudes, the non-orthogonality usually brings interesting effects when energy damping is present \cite{semiao}.  The use of coherent states in quantum information is in fact a very active research topic giving that the generation, manipulation and characterization of these states are very well established in experimental quantum optics \cite{book_exp_qo}. The use of coherent states for quantum teleportation and quantum logic is detailed in \cite{jeong2,jeong1,milburn,semiao2}. Entangled coherent states, such as the ones used in this work, have in fact been experimentally generated in superconducting circuits  where full quantum state tomography have been successfully implemented \cite{two_boxes}. In general, one needs some nonlinearity in the Hamiltonian to dynamically superpose or entangle coherent states \cite{c1,Sanders}.

This work is organized as follows. In Sec.~\ref{be}, we quickly review the CT protocol and present our investigation problem which is the performance of different optical qubits in controlled teleportation under dissipation. Sec.~\ref{res} is dedicated to the presentation of our results. In particular, we compare the performance of single-rail encoding using orthogonal Fock states with that using coherent states. We also look into the tripartite non-locality of the state shared by the parties to see whether or not non-locality and efficiency are related. In Sec. \ref{conc}, we sumarize our findings and present our final remarks. Finally, in the Appendix, we briefly review a method to evaluate teleportation fidelities.
\section{Basic elements}\label{be}
\subsection{Controlled teleportation protocol}
In the context of CT, we will assume that the controller (Charlie) possesses qubit $1$, while the sender (Alice) and the receiver (Bob) possess qubits $2$ and $3$, respectively. In the ideal case (no dissipation), these three qubits are in the pure MS state defined as \cite{ghose1,msstate}
\begin{align}\label{MS1}
\ket{MS}_{123}=\frac{1}{\sqrt{2}}(\ket{0,0,0}+c\ket{1,1,1}+d\ket{0,1,1})_{123},
\end{align}
where $c$ and $d$ are real numbers subjected to $c^2=1-d^2$. Actually, in the simulations, we will be using $c=\cos\theta$ and $d=\sin\theta$ with $\theta\in[0,\pi/2]$. Alice and Bob will use their part of this state to perform the standard teleportation \cite{tp} of the unknown state of a forth qubit in the possession of Alice.  To see how Charlie can affect the performance of the teleportation protocol, we rewrite Eq.(\ref{MS1}) as
\begin{align}\label{MSa}
\ket{MS}_{123}=&\frac{1}{2}[(1+d)\ket{0}+c\ket{1}]_{1}\otimes\ket{\Phi^+}_{23}\nonumber\\+&\frac{1}{2}[(1-d)\ket{0}-c\ket{1}]_{1}\otimes\ket{\Phi^-}_{23},
\end{align}
where $\ket{\Phi^\pm}_{23}=1/\sqrt{2}(\ket{0,0}\pm\ket{1,1})_{23}$ are maximally entangled Bell states. Now, it is easy to see that if Charlie performs a projective measurement on the orthonomal basis  \cite{ghose1}
\begin{align}\label{base}
\ket{\xi^\pm}_1=\frac{1}{\sqrt{(1\pm d)^2+c^2}}[(1\pm d)\ket{0}\pm c\ket{1}]_1,
\end{align}
Alice and Bob end up with one of two Bell states. Upon receiving a classical message from Charlie announcing his measurement output, they can perform the teleportation with unit fidelity. The teleportation fidelity under these conditions will be called conditioned fidelity  $F_c$. The local measurement by Charlie and subsequent announcement of the output to Alice and Bob constitute the \textit{controller permission}. 

In the controlled teleportation protocol, it is demanded that Alice and Bob are at least semi-honest, i.e., they may want to perform teleportation without Charlie's permission, but they will not establish their on bipartite quantum channel. In other words, they follow the protocol but may try to cheat the controller \cite{shtg}. There is a great deal of papers dealing with different facets of the concept of semi-honest parties and exploring their consequences in varied in quantum protocols \cite{sh}. When Charlie does not give the permission, it is the non-conditioned teleportation fidelity $F_{nc}$ that should be investigated. It is evaluated with the corresponding density operator for the MS state in Eq.(\ref{MSa}), after the controller has been traced out. This corresponds to a scenario where no measurements were performed by the controller. In this case, one finds \cite{ghose1}
\begin{align}
\label{fnc1}
F_{nc}=\frac{2+d}{3}.
\end{align}
From this result, it is clear that for $d=0 (\theta=0)$, when the MS state in Eq.(\ref{MS1}) reduces to the GHZ state, Alice and Bob fails completely in the attempt of cheating \cite{ctfirst}. In this case, all they obtain is the classical fidelity $2/3$.  As $d$ increases, the power of the controller diminishes. In the limit case $d=1(\theta=\pi/2)$, the teleportation is successfully performed without permission of the controller. Again, this is a simple consequence of form of the MS state in Eq.(\ref{MS1}). When $d=1$,  Alice and Bob are left in a Bell state which is completely uncorrelated with the qubit with Charlie.

In \cite{ghose1}, $1-F_{nc}$ is called control power. Here, we will renormalize it in a way that it explicitly takes into account the unsuitability of fidelities below $2/3$ for quantum teleportation. This normalized control power $C_p$ varying between zero and one reads
\begin{align}
\label{cp}
C_p=\left\{
\begin{array}{c}
1-3\left(F_{nc}-\frac{2}{3}\right),\ \ {\rm{if}} \ \ F_{nc}>2/3\\
1, \ \ {\rm{if}} \ \ F_{nc}\leq 2/3
\end{array},
\right.
\end{align}
Notice that $F_{nc}=1$ implies $C_p=0$. On the other hand, fidelities equal or below the classical bound imply $C_p=1$. These two limits correspond to complete success or fail of Alice and Bob when they proceed without permission of Charlie.

The CT protocol is intrinsically a twofold problem.  First, when no permission is given, $C_p$ must be as close to one as possible. Second, when the permission is given, the conditioned fidelity $F_c$ must be higher than $2/3$ and as close as possible to $1$.  Consequently, the product of the control power and the conditioned fidelity provides us with information about the performance of the controller or efficiency of the protocol. Upon normalization of the conditioned fidelity, i.e., $F_c\rightarrow [1+3(F_c-1)]$, meaning that   zero is attributed to the classical bound $(F_c=2/3)$ and one to $F_c=1$, we then define an efficiency $\eta$ for the CT protocol as
\begin{align}
\label{eficiencia}
\eta=\left\{
\begin{array}{c}
C_p[1+3(F_c-1)],\ \ {\rm{if}} \ \ F_c>2/3\\
0, \ \ {\rm{if}} \ \ F_c\leq 2/3
\end{array},
\right.
\end{align}
Notice that $\eta$ varies between zero and one and, whenever $F_c\leq2/3$ or $C_p=0$, the efficiency is null. This means that, in order for the control to be effective, it is of no use to have a high control power $C_p$ when the conditioned teleportation fidelity $F_c$ is under the classical bound. 

When defining an efficiency or quantifying a resource, one always have in mind an specific scenario. For the reasons stated above, the product of conditioned fidelity and control power is very appropriate for our purposes. Of course, given another scenario, the choice might change. For instance, one could think of defining an efficiency as the difference between the fidelities with and without controller permission: $\max[F_c-F_{nc},0]$. In this case, one is interested in a scenario where a failure is atribute to the protocol (controller) when Alice and Bob achieve better fidelities with no permission than with permission. Of course, this is also a legitimate standpoint and the results are sensitive to the scenario one wants to investigate. From our standpoint, we want to atribute a finite efficiency when $F_c<F_{nc}$, as long as $C_p$ is not null. The normalized product defined in Eq.(\ref{eficiencia}) clearly performs well in this scenario. We now proceed to apply these quantifiers $F_c$, $C_p$, and $\eta$ in the study of the performance of different optical qubits in CT with MS states and damping.
\subsection{The problem}
\begin{figure}[!h]
	\includegraphics[width=0.7\linewidth]{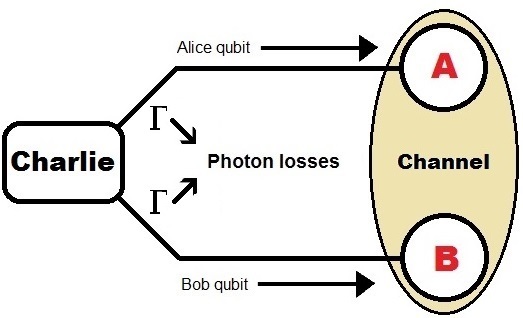}
	\vspace{-0.2cm}
	\caption{(Color online) Charlie prepares the MS state in Eq.(\ref{MS1}), send one optical qubit to Alice and another to Bob, both through optical fibers with dissipation rate $\Gamma$.  During this propagation, the qubit that stays with Charlie is assumed to approximately not dissipate energy nor suffer decoherence (closed system).}
	\label{esquema}
\end{figure}

The physical situation we want to address in this work is depicted in Fig.\ref{esquema}. 
The key figure in the CT protocol is Charlie, the controller.  It is then natural to think that it is him who will prepare the tripartite MS state and distribute one qubit to Alice and another to Bob. In his laboratory, Charlie keep a long-lived stationary matter qubit based, for instance, on electronic degree of freedom of trapped ions or states of artificial atoms in circuit quantum electrodynamics. However, for Alice and Bob, Charlie must employ flying or optical qubits that will be sent to their laboratories. Typically, this is made with the help of optical fibers which inevitably involve losses (amplitude damping). Time evolution under amplitude damping is described by the master equation \cite{louisell}
\begin{align}
\frac{\partial \rho}{\partial t}=J\rho+L\rho,
\label{BMeq}
\end{align}
where $J\rho=\Gamma\sum_{i=2}^{3} a_i\rho a_i^\dagger$ and $L\rho=-(\Gamma/2)\sum_{i=2}^{3} (a_i^\dagger a_i\rho + \rho a_i^\dagger a_i)$ with $a_i$($a_i^\dagger$) the annihilation (creation) operator acting on mode $i$, and $\Gamma$ is the dissipation rate of the optical fibers. Notice that, in our problem, only the modes addressed to Alice and Bob $(i=2,3)$ are subjected to damping in the fibers. In the literature, there are other versions of the controlled teleportation protocol which are intrinsically probabilistic. This type of protocol will not be considered here, and its performance with decoherence is studied in \cite{Wct}.  

As mentioned before, we will be working with two distinct physical qubits in single-rail logic. They are
\begin{enumerate}
	\item {\it Vacuum and Single-Photon States} (VSP): in this case, the physical qubits are the Fock states corresponding to zero or one photon in the mode, i.e., 
\begin{eqnarray}\label{k1}
	\{\ket{0},\ket{1}\}\longrightarrow\{\ket{\bar{0}},\ket{\bar{1}}\}.
\end{eqnarray}
The bars are used in the VSP  to avoid confusion between physical and logical qubits. Using the master equation in Eq.(\ref{BMeq}) one finds
\begin{eqnarray}
\ket{\bar{0}}\bra{\bar{0}}&\rightarrow&\ket{\bar{0}}\bra{\bar{0}}\nonumber\\
\ket{\bar{1}}\bra{\bar{0}}&\rightarrow&\tau^2\ket{\bar{1}}\bra{\bar{0}}\nonumber\\
\ket{\bar{1}}\bra{\bar{1}}&\rightarrow&\tau^2\ket{\bar{1}}\bra{\bar{1}}+r^2\ket{\bar{0}}\bra{\bar{0}},
\end{eqnarray}
where $\tau=e^{-\Gamma t/2}$ and $r=\sqrt{1-\tau^2}$ is a normalized time varying from zero to one. 

	\item {\it Coherent States}: coherent states $\ket{\alpha}$ in Fock basis is defined as  $\ket{\alpha}=e^{-|\alpha|^2/2}\sum_{n=0}^{\infty}(\alpha^n/\sqrt{n!})\ket{n}$ with $\alpha$ complex. For the encoding, we will be employing coherent states of same amplitude but opposite phases  \cite{jeong2,jeong1,milburn,semiao2}, i.e.,
\begin{eqnarray}  \label{k2}
	\{\ket{0},\ket{1}\}\longrightarrow\{\ket{\alpha},\ket{-\alpha}\}.
\end{eqnarray} 	
By using again the master equation in Eq.(\ref{BMeq}) one finds now
\begin{eqnarray}
\ket{\alpha}\bra{\alpha}&\rightarrow&\ket{\gamma}\bra{\gamma}\nonumber\\
\ket{\alpha}\bra{-\alpha}&\rightarrow&e^{-2r^2|\alpha|^2}\ket{\gamma}\bra{-\gamma},
\end{eqnarray}
where  $\gamma=\alpha \tau$.
\end{enumerate}
\section{Results}\label{res}
For each of the photonic qubits considered in this work, and the initial MS state in Eq.(\ref{MS1}), we now study how the conditioned fidelities and the control power are affected by losses in the fibers \cite{jeong3}. After individually studying these figures of merit, we will then directly compare the encodings using the efficiency as defined in Eq.(\ref{eficiencia}). Finally, we will investigate whether there is any meaningful relation between tripartite non-locality and the efficiency of the protocol.
\subsection{Vacuum and single-photon}
The first step is to rewrite the MS state in Eq.(\ref{MS1}) in terms of the physical qubits of the VSP encoding in Eq.(\ref{k1}). Then, Eq.(\ref{BMeq}) is solved with the initial condition $\rho(0)=\ket{MS}_{123}\bra{MS}$. In the case of the non-conditioned density operator for Alice and Bob,  qubit 1 (controller qubit) is traced out resulting in
\begin{eqnarray}	
\rho_{nc}(t)&=&\frac{1}{2}[(1+r^4)\ketbra{\bar{0},\bar{0}}{\bar{0},\bar{0}}+\tau^4\ketbra{\bar{1},\bar{1}}{\bar{1},\bar{1}}\nonumber\\ &&+\tau^2\sin\theta(\ketbra{\bar{0},\bar{0}}{\bar{1},\bar{1}}+\ketbra{\bar{1},\bar{1}}{\bar{0},\bar{0}})\nonumber\\
&&+r^2\tau^2(\ketbra{\bar{0},\bar{1}}{\bar{0},\bar{1}}+\ketbra{\bar{1},\bar{0}}{\bar{1},\bar{0}})],
\label{rovspnc}
\end{eqnarray}

For the conditioned fidelity $F_c$, instead of tracing out qubit 1, the evolved density operator is projected onto the basis in Eq.(\ref{base}). The two possible outcomes are
\begin{eqnarray}
\rho_c^\pm(t)&=&\frac{1}{2}[(1+r^4)\ketbra{\bar{0},\bar{0}}{\bar{0},\bar{0}}+\tau^4\ketbra{\bar{1},\bar{1}}{\bar{1},\bar{1}}\nonumber\\ &&\pm\tau^2(\ketbra{\bar{0},\bar{0}}{\bar{1},\bar{1}}+\ketbra{\bar{1},\bar{1}}{\bar{0},\bar{0}})\nonumber\\
&&+r^2\tau^2(\ketbra{\bar{0},\bar{1}}{\bar{0},\bar{1}}+\ketbra{\bar{1},\bar{0}}{\bar{1},\bar{0}})],
\label{rovspc}
\end{eqnarray}
which show up with the probabilities
\begin{align}
P^\pm=\frac{1\pm\sin\theta}{2}.
\label{vspprob}
\end{align}
Now, the teleportation fidelities 
\begin{eqnarray}
F_c&=&P^+F(\rho_c^+)+P^-F(\rho_c^-),\label{dc}\\ 
F_{cn}&=&F(\rho_{nc})
\end{eqnarray}
can be evaluated using the method discussed in the Appendix. The result is
\begin{eqnarray}
F_{nc}&=& \frac{1}{6}(3+2 \sin\theta |r^2-1|+|1-2r^2+2r^4|),\label{f1}\\
F_c&=&\frac{1}{6}(3+2 |r^2-1|+|1-2r^2+2r^4|).\label{f2}
\end{eqnarray}
Please, notice that the conditioned fidelity  $F_c$ defined in Eq.(\ref{dc}) is actually an average fidelity weighted by the probabilities of obtaining each of the conditioned density operators upon the measurement performed by Charlie. For the particular case of the VSP qubits, these probabilities are given by Eq.(\ref{vspprob}). For the qubits based on coherent states to be discussed next, these probabilities will naturally change. 

\begin{figure}[!htb]
		\begin{minipage}[b]{0.49\linewidth}
			\includegraphics[width=\linewidth]{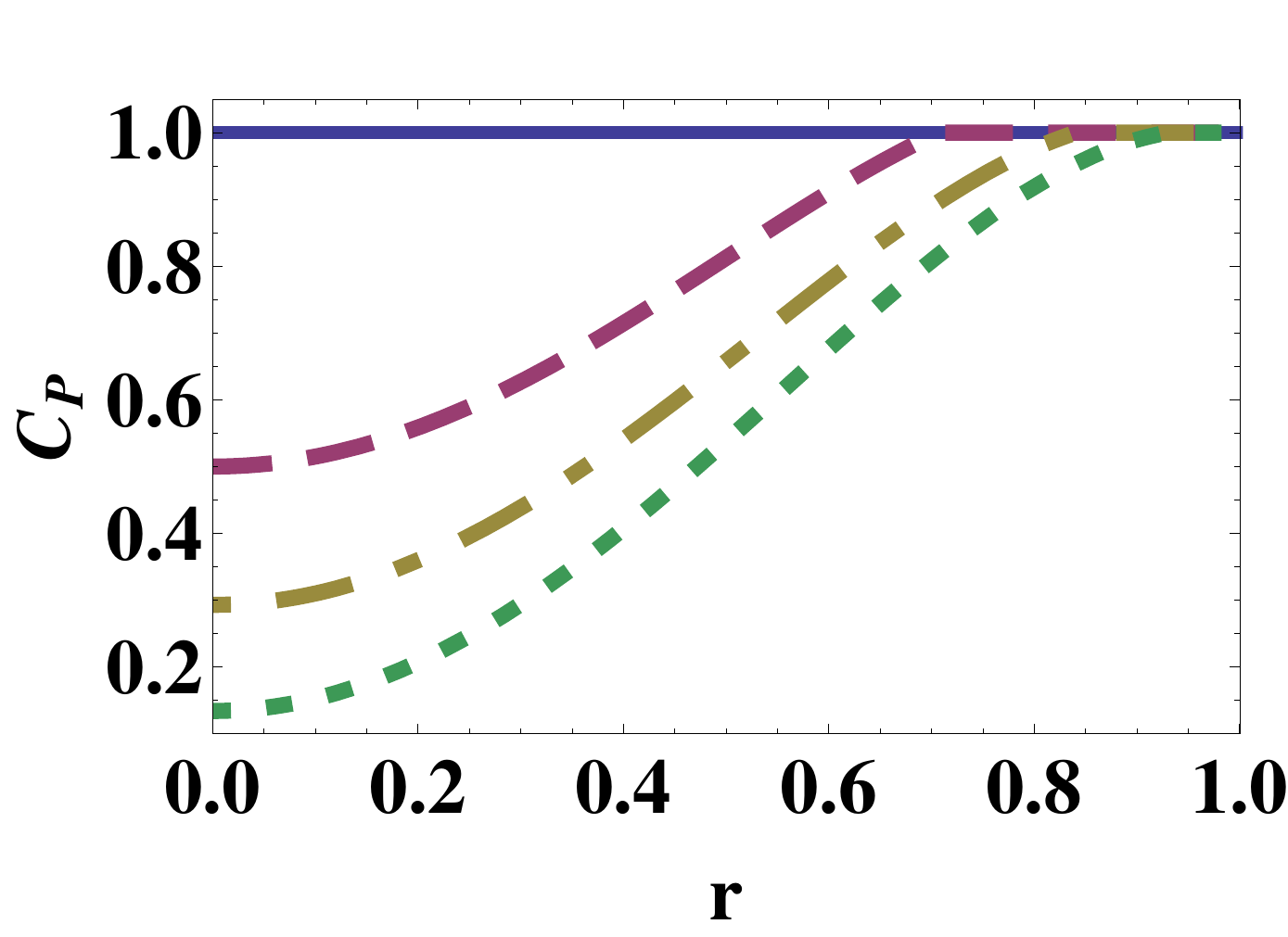}
		\end{minipage} 
		\begin{minipage}[b]{0.49\linewidth}
			\includegraphics[width=\linewidth]{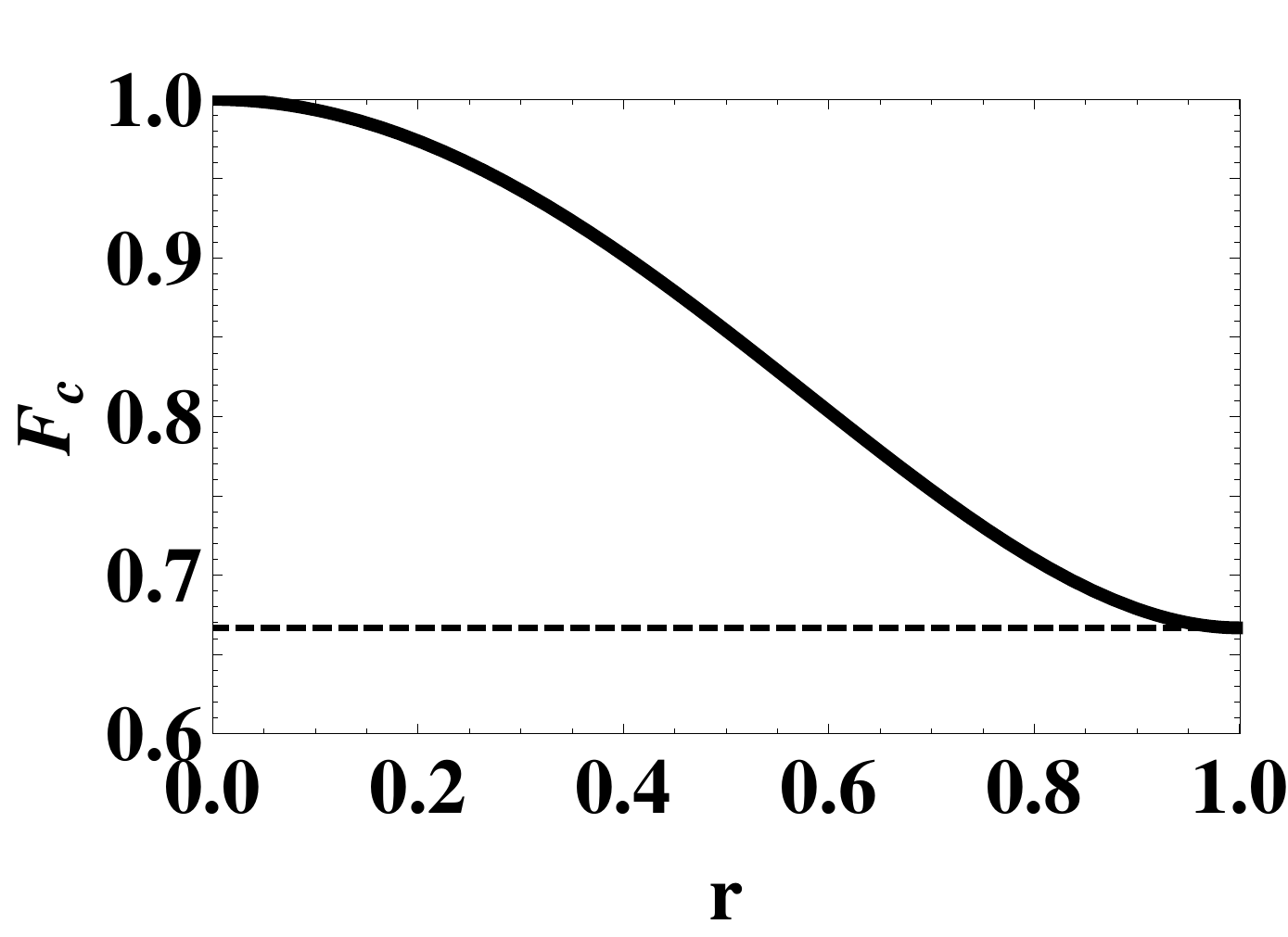}
		\end{minipage}\hspace{0cm}
		\vspace{-0.2cm}
		\caption{(Color online) Control power (left) and conditioned fidelity (right) as a function of normalized time and different values of $\theta$ for the VSP encoding. Left panel: $\theta=0$ (solid), $\theta=\pi/6$ (dashed), $\theta=\pi/4$ (dot-dashed), and $\theta=\pi/3$ (dotted). Right panel: $F_c$ turned out to be independent on $\theta$ [see Eq.\ref{f2}] and the dashed straight line indicates de classical fidelity $2/3$.}
		\label{VSP}
	\end{figure}

In Fig. \ref{VSP}, we present the effect of losses in the fiber for the control power evaluated with Eqs. (\ref{cp}) and (\ref{f1}) and the conditioned fidelity (\ref{f2}) for different values of $\theta$, i.e., some choices of $c$ and $d$ in the MS state Eq.(\ref{MS1}). First, it is worth noticing that for $\theta=0$ the control power remains one regardless of the losses in the fiber. This is so because $\theta=0$ corresponds to a maximally entangled GHZ state, whose reduced state (tracing out the controller) is a separable state. Since the local dissipative dynamics is unable to create entanglement,  Alice and Bob will always fail to perform quantum teleportation. On the other hand, as $\theta$ is progressively increased in the range $[0,\pi/2]$, the initial state with Alice and Bob tends continually to a maximally entangled Bell state, causing $C_p$ to decrease. In this context, the losses in the fibers seem to help the controller (increasing $C_p$) as they destroy the entanglement in the reduced state of Alice and Bob. This is where the introduction of the efficiency quantifier in Eq.(\ref{eficiencia}) will play an important role. The presence of losses will also affect the conditioned fidelity $F_c$ (see  Fig. \ref{VSP},), obtained with the permission of the controller. Consequently, the increment of $C_p$ due to the losses are not necessarily good for CT protocol. 
\subsection{Coherent state}
We now proceed to the second kind of photonic qubit considered in this work. Now, the first step is to rewrite the MS state in Eq.(\ref{MS1}) in terms of the  encoding in Eq.(\ref{k2}). It is also necessary to renormalize the state since the coherent states used in the encoding are not orthogonal. Next, Eq.(\ref{BMeq}) is solved with the initial condition $\rho(0)=\ket{MS}_{123}\bra{MS}$. After tracing out  qubit 1 (controller qubit), we get the non-conditioned density operator
\begin{eqnarray}
\rho_{nc}(t)&=&N(\alpha)^2[\ketbra{\gamma,\gamma}{\gamma,\gamma}+\ketbra{-\gamma,-\gamma}{-\gamma,-\gamma}\nonumber\\ &&+e^{-4r^2|\alpha|^2}\sin\theta \ketbra{\gamma,\gamma}{-\gamma,-\gamma}+\nonumber\\
&&+ e^{-4r^2|\alpha|^2}\sin\theta\ketbra{-\gamma,-\gamma}{\gamma,\gamma}],
\end{eqnarray}
where  $N(\alpha)=1/\sqrt{2+2 \sin\theta\exp[-4|\alpha|^2]}$. For the conditioned case, we again consider projection upon the basis in Eq.(\ref{base}), and the two possible outcomes are now
\begin{eqnarray}
\rho_c^\pm(t)&=&M_\pm(\alpha)^2[\ketbra{\gamma,\gamma}{\gamma,\gamma}+\ketbra{-\gamma,-\gamma}{-\gamma,-\gamma}\nonumber\\ &&\pm e^{-4r^2|\alpha|^2}\ketbra{\gamma,\gamma}{-\gamma,-\gamma}\nonumber\\ &&\pm e^{-4r^2|\alpha|^2}\ketbra{-\gamma,-\gamma}{\gamma,\gamma}],
\label{rocsc}
\end{eqnarray}
where  $M_\pm(\alpha)=1/\sqrt{2\pm2\exp[-4|\alpha|^2]}$. The probabilities for $\rho_c^\pm(t)$ are
\begin{align}
P_\alpha{}^\pm=\frac{1\pm \sin\theta}{2(1+\sin\theta e^{-4|\alpha|^2})}(1\pm e^{-4|\alpha|^2}).
\label{csprob}
\end{align}
These probabilities will be used to evaluate the conditioned fidelity $F_c=P_\alpha{}^+F(\rho_c^+)+P_\alpha{}^-F(\rho_c^-)$, as explained before. In order to once again use the methods discussed in the Appendix, we need to redefine a proper magic basis in terms of coherent states. It now reads
\begin{align}
&\ket{m_1^\gamma}=\frac{1}{\sqrt{2}}(\ket{\gamma^+,\gamma^+}+\ket{\gamma^-,\gamma^-}),\nonumber\\
&\ket{m_2^\gamma}=\frac{i}{\sqrt{2}}(\ket{\gamma^+,\gamma^+}-\ket{\gamma^-,\gamma^-}),\nonumber\\
&\ket{m_3^\gamma}=\frac{i}{\sqrt{2}}(\ket{\gamma^+,\gamma^-}+\ket{\gamma^-,\gamma^+}),\nonumber\\
&\ket{m_4^\gamma}=\frac{1}{\sqrt{2}}(\ket{\gamma^+,\gamma^-}-\ket{\gamma^-,\gamma^+}), \nonumber
\end{align}
where $\ket{\gamma^\pm}=(2\pm2e^{-2|\gamma|^2})^{-1/2}(\ket{\gamma}\pm\ket{-\gamma})$ are even and odd coherent states which are naturally orthonormal $\braket{\gamma^+}{\gamma^-}=0$. The expressions for $F_c$ and $C_p$ are now a bit cumbersome, so we will only perform a graphic analysis of these quantities.
\begin{figure}[htb!] 
		\begin{minipage}[b]{0.49\linewidth}
			\includegraphics[width=\linewidth]{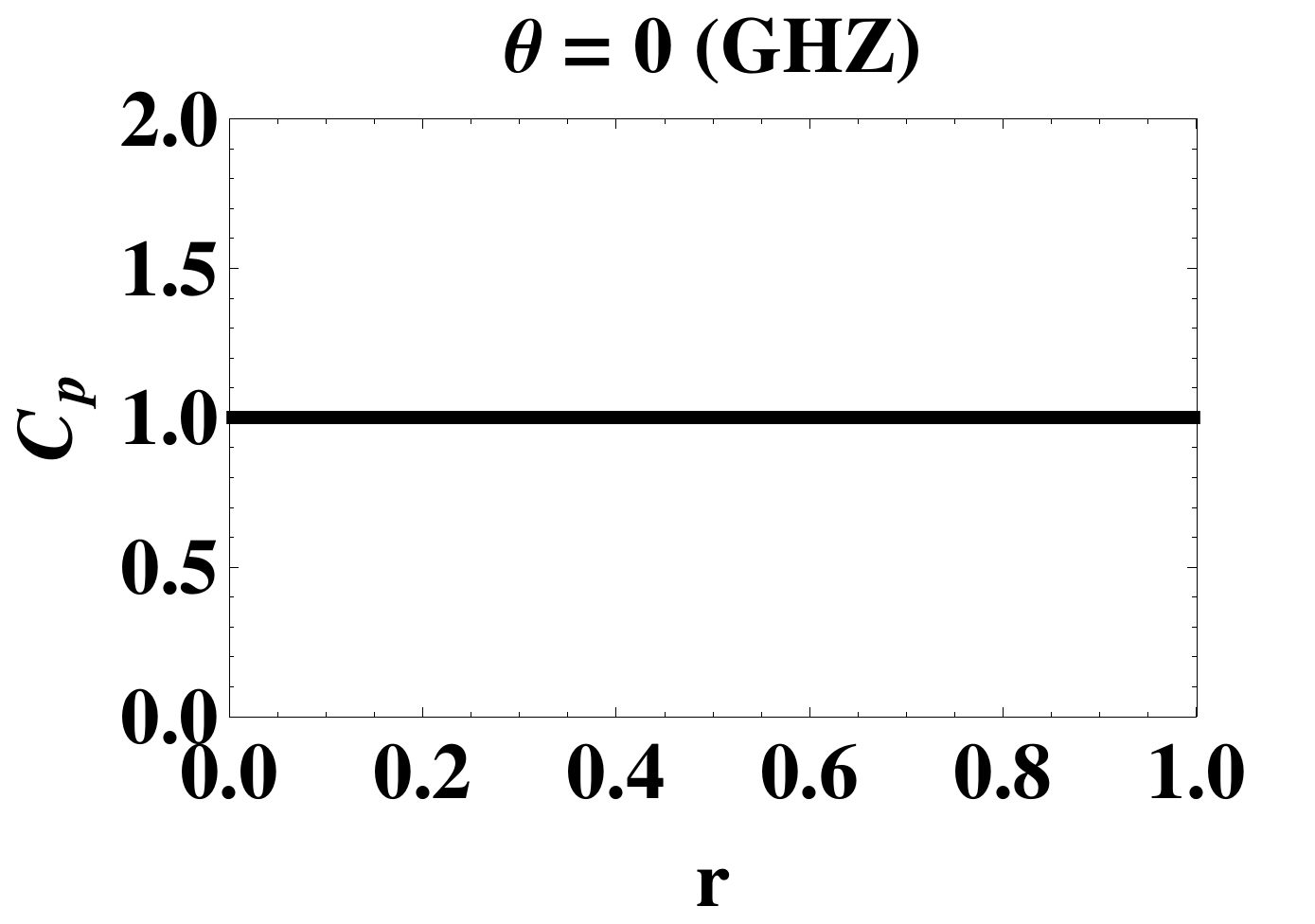}
		\end{minipage}
		\begin{minipage}[b]{0.49\linewidth}
			\includegraphics[width=\linewidth]{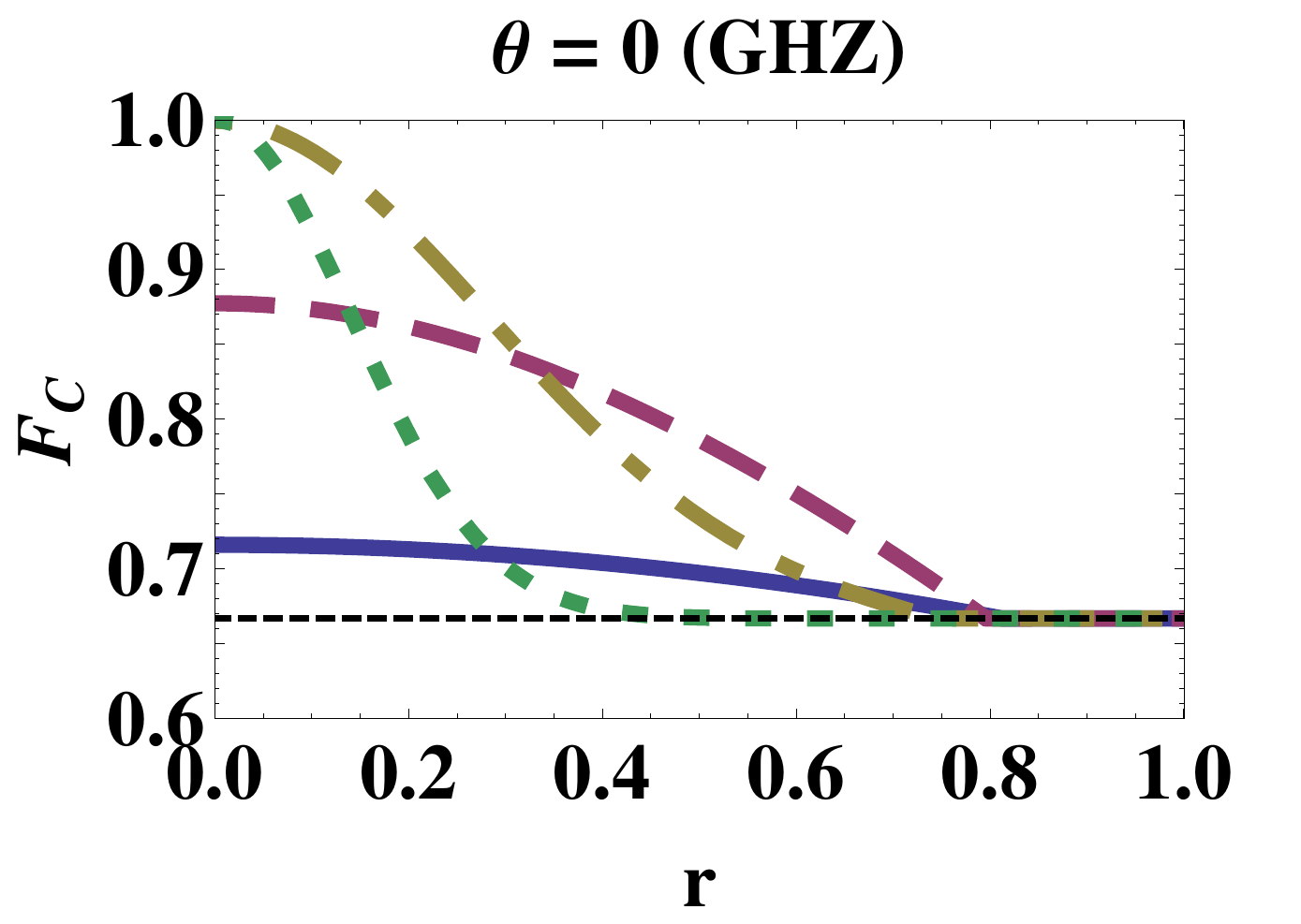}
		\end{minipage}
		\begin{minipage}[b]{0.49\linewidth}
			\includegraphics[width=\linewidth]{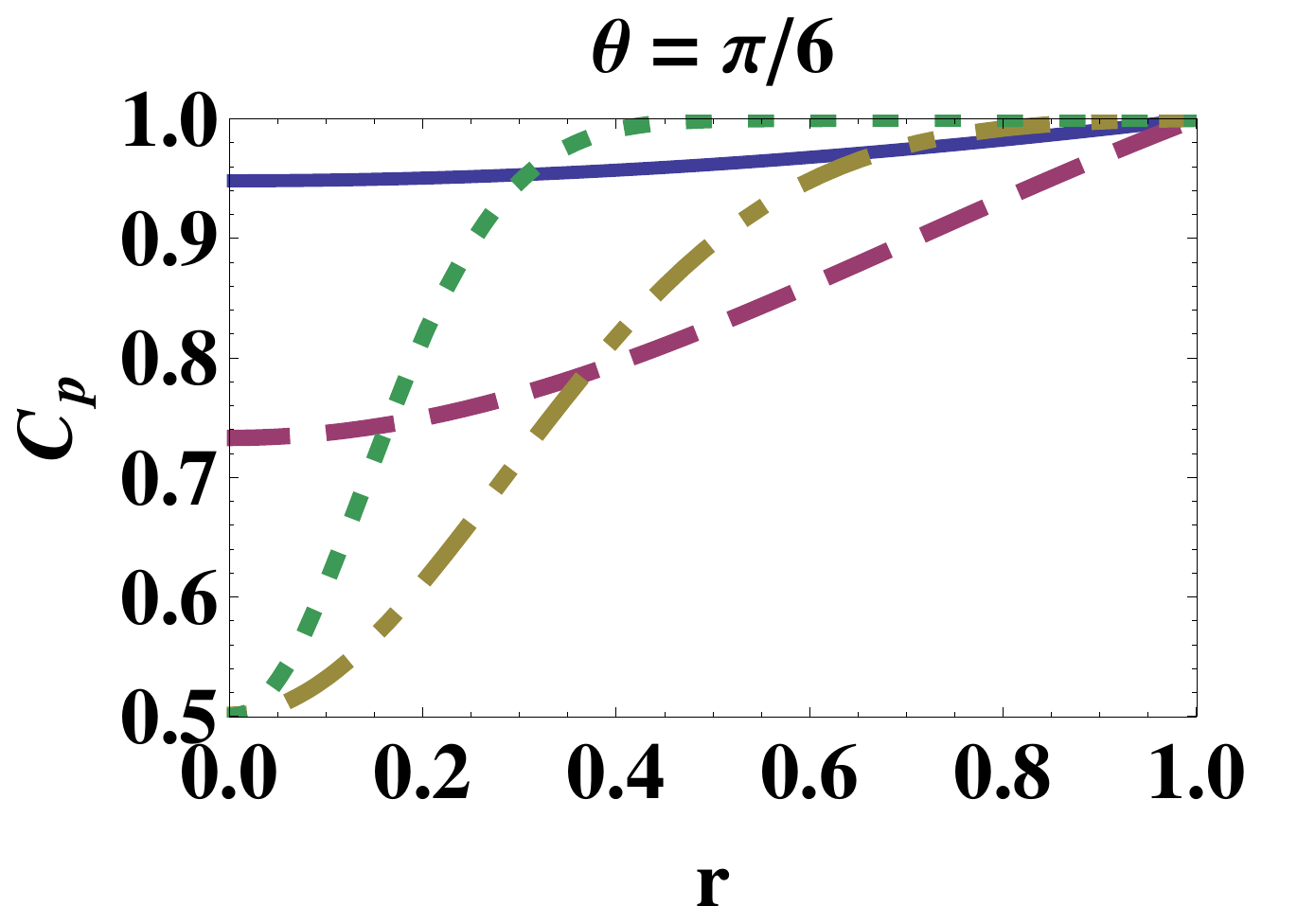}
		\end{minipage}
		\begin{minipage}[b]{0.49\linewidth}
			\includegraphics[width=\linewidth]{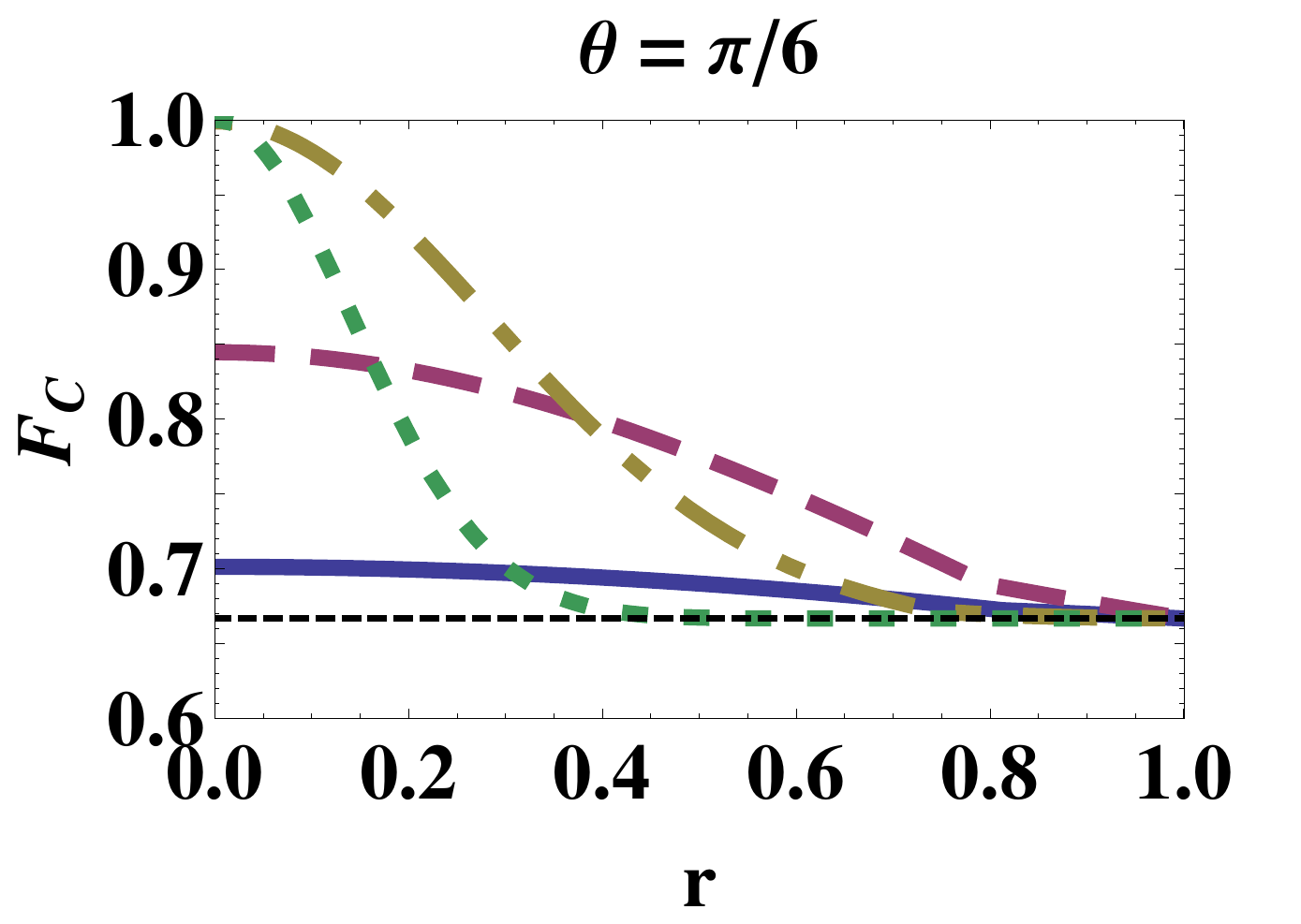}
		\end{minipage}
		\begin{minipage}[b]{0.49\linewidth}
			\includegraphics[width=\linewidth]{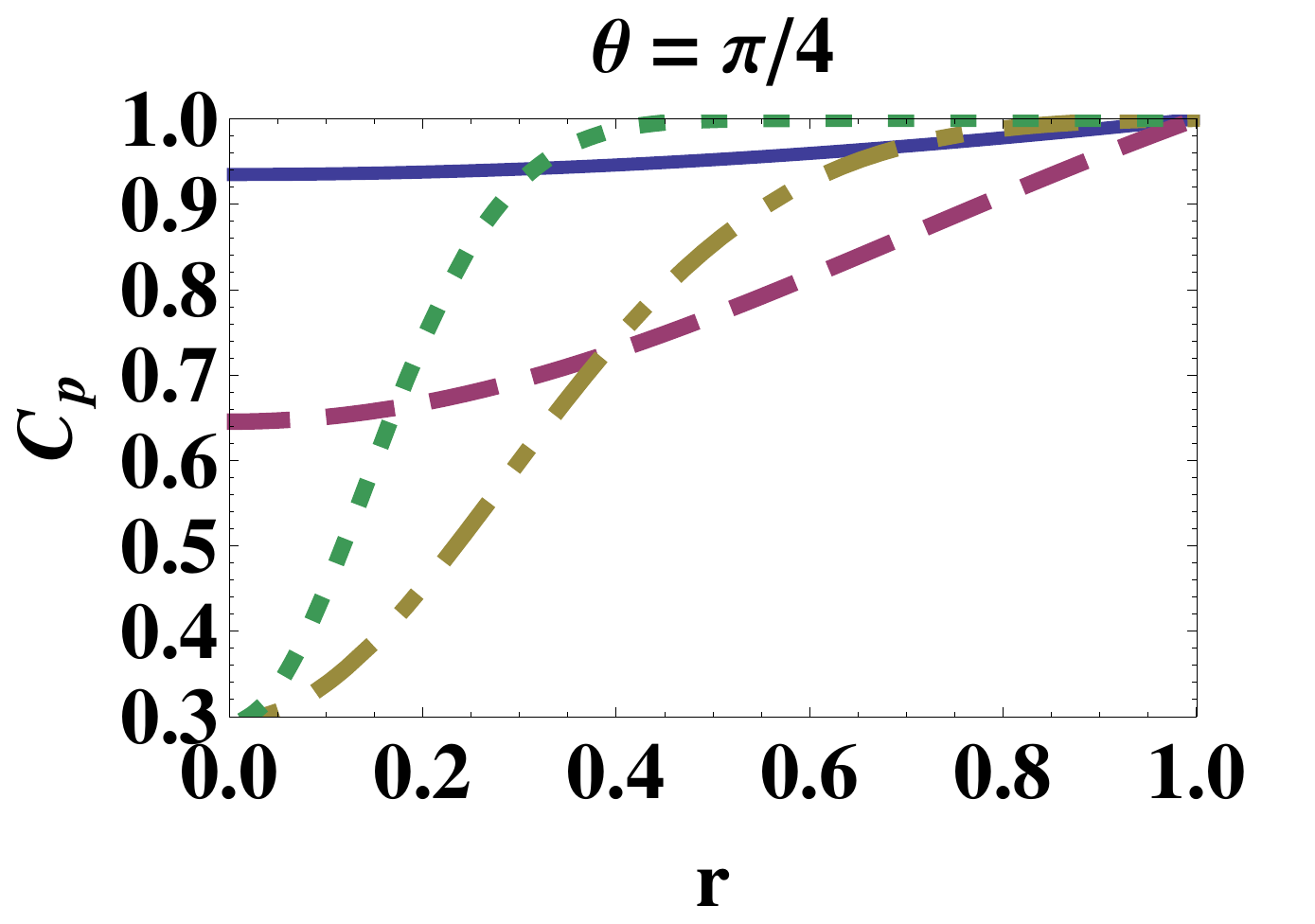}
		\end{minipage}
		\begin{minipage}[b]{0.49\linewidth}
			\includegraphics[width=\linewidth]{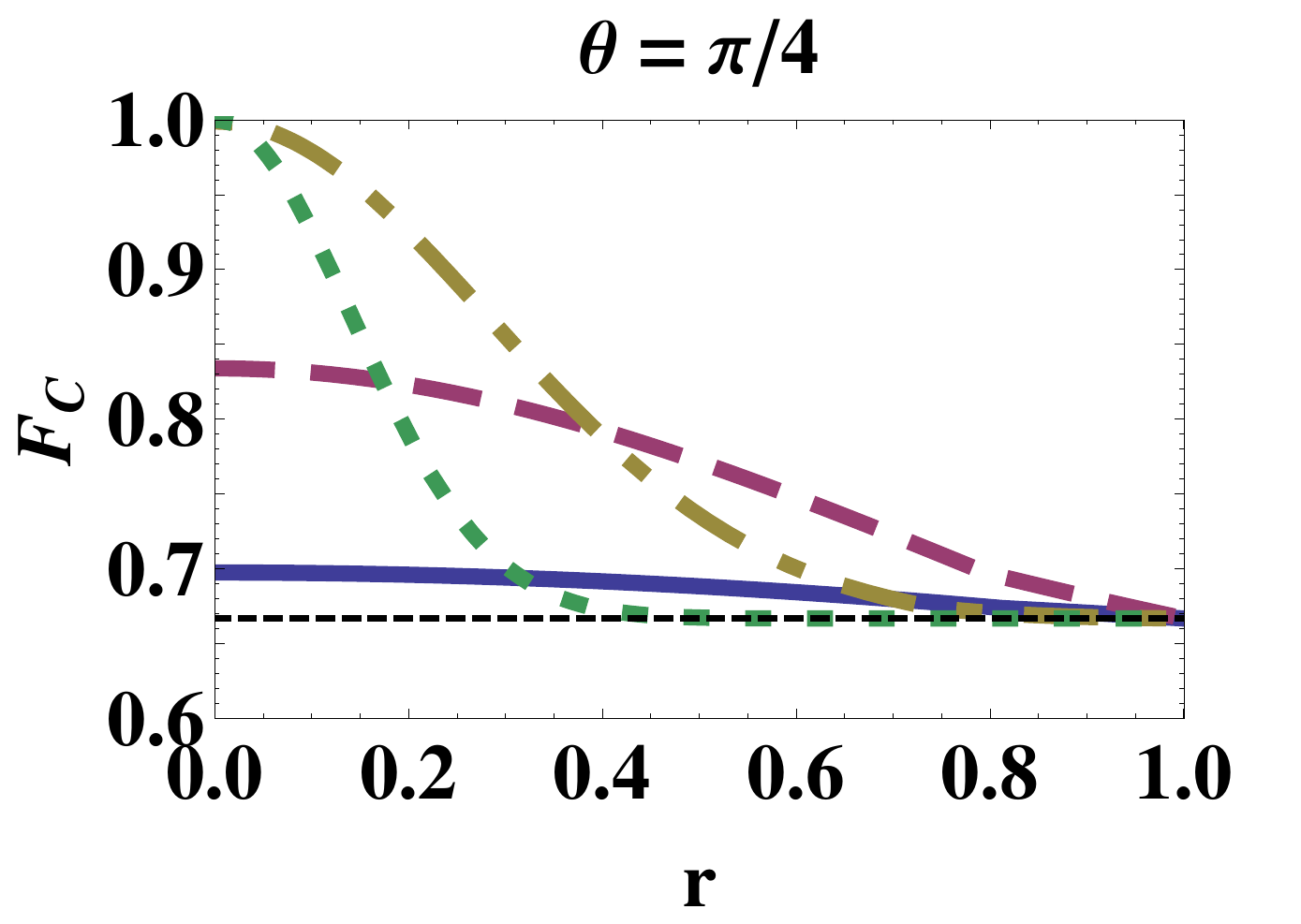}
		\end{minipage}
		\begin{minipage}[b]{0.49\linewidth}
			\includegraphics[width=\linewidth]{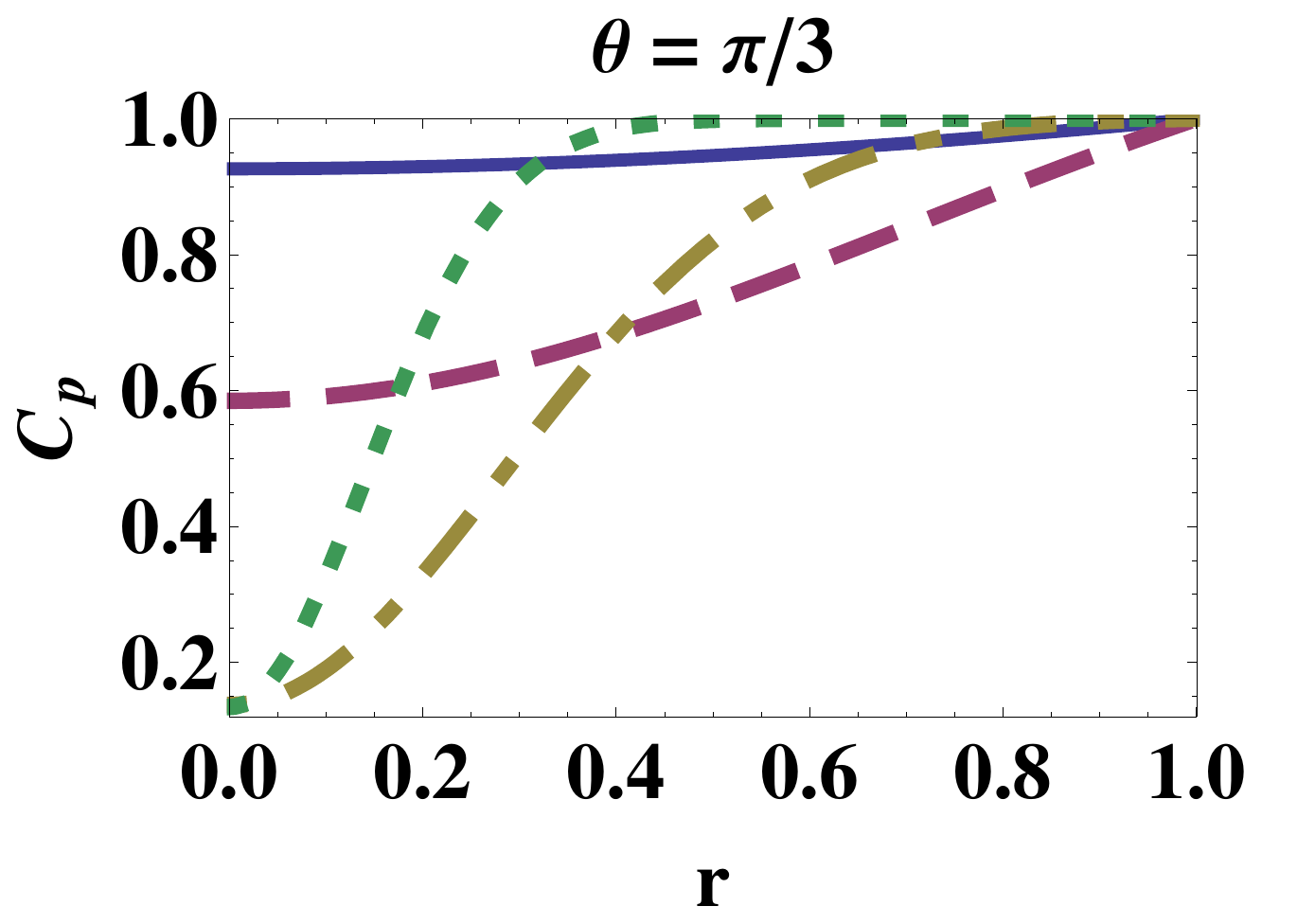}
		\end{minipage}
		\begin{minipage}[b]{0.49\linewidth}
			\includegraphics[width=\linewidth]{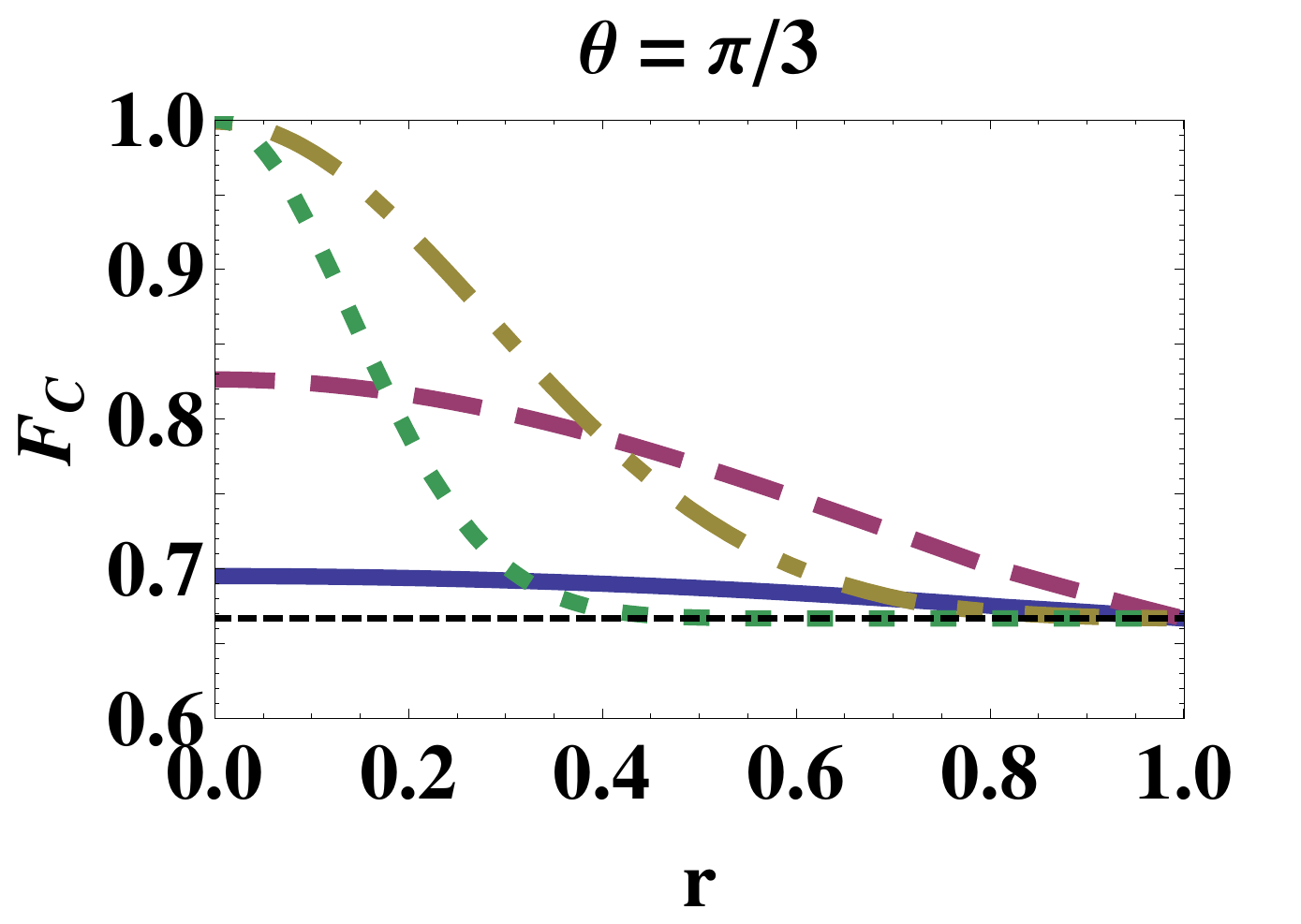}
		\end{minipage}
		 \vspace{-0.2cm}
		\caption{(Color online) Control power (left) and conditioned fidelity (right) as a function of normalized time and different values of $\theta$ and $\alpha$ for the coherent states encoding. The dashed straight line indicates the classical fidelity $2/3$. The used amplitudes of the coherent states are $\alpha=0.20$ (solid), $\alpha=0.50$ (dashed), $\alpha=1.25$ (dot-dashed), $\alpha=2.50$ (dotted).}
		\label{CS}
\end{figure}
	
In Fig.\ref{CS}, we present the plots of the control power and conditioned fidelity for different values of $\theta$ e $\alpha$. When the initial state corresponds to a type of GHZ ($\theta=0$), the control power, besides being independent of $r$,  becomes now also independent of $\alpha$. Once again this happens because the reduced state of Alice and Bob possess no entanglement for $\theta=0$. This is not observed for $F_c$ which presents a strong dependence on $\alpha$ even when $\theta=0$. In particular, we want to remark the existence of different crossings when $r$ is varied.  This means that, depending on the normalized time i.e., the losses in the fiber, it is more appropriate to choose large or small coherent state amplitudes $\alpha$ to achieve higher control or conditioned fidelities. This happens for two reasons. First, superposition of coherent states are more affected by the losses when their amplitudes are large \cite{decoerence}. However, large amplitudes  produce better conditioned fidelities because the states in Eq.(\ref{k2}) tend to orthogonality. It is this compromise between dissipation and orthogonality which produces the crossings observed in Fig.\ref{CS} for intermediate values of $r$. All these make the coherent states interesting from the quantum information perspective  \cite{semiao,pj}. As a final remark, at $r=0$, the control power decreased with $\theta$ because, in the limit $\theta\rightarrow\pi/2$, the reduced state is a quasi-Bell state
\begin{align}\label{qB}
&\ket{\Phi^+_\alpha}=(2+2e^{-4|\alpha|^2})^{-1/2}(\ket{\alpha,\alpha}+\ket{-\alpha,-\alpha})
\end{align}
 which allows almost perfect teleportation with large $\alpha$ \cite{semiao}.
\subsection{Efficiency}
From the previous discussions, it is clear that the control power $C_p$ and the conditioned fidelities $F_c$, usually employed in the analysis of CT protocol \cite{ghose1}, are not enough to grasp all aspects of the problem when dissipation is present or non-orthogonal  states are used in the encoding. In particular, the plots in  Fig.\ref{CS} clearly show that, for small $r$ (low dissipation), an increase in $\alpha$ can help $F_c$, what is desirable in the CT protocol. However, this causes $C_p$ to decrease and this is not good for the protocol. This situation becomes even more evolved for moderate dissipation where crossings do happen. It is for this reason that we now proceed to analyze the efficiency $\eta$ as defined in Eq.(\ref{eficiencia}). This quantity will allow us to finally compare the performance of the two different optical qubits.

\begin{figure}[!htb]
	\begin{minipage}[b]{0.49\linewidth}
		\includegraphics[width=\linewidth]{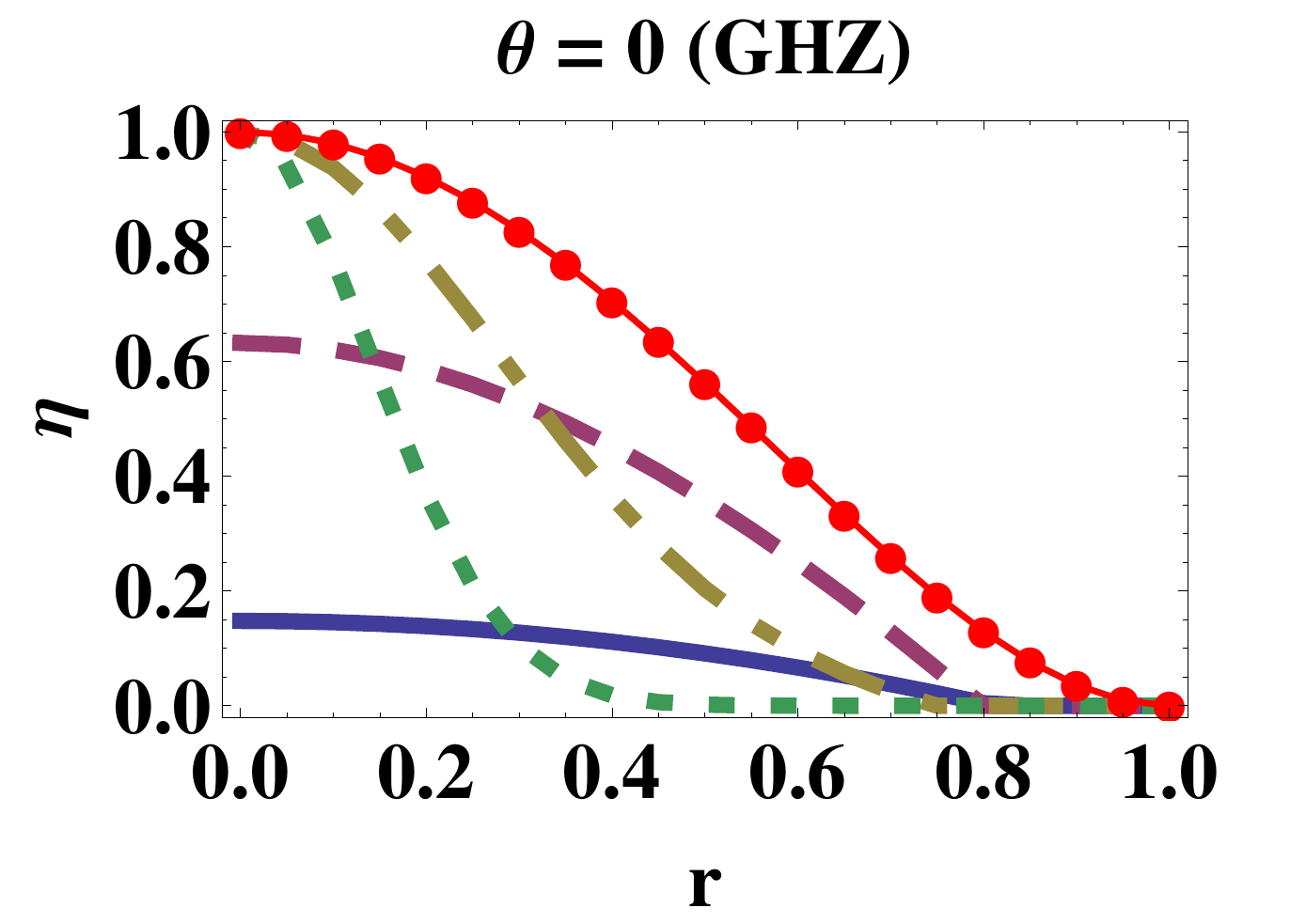}
	\end{minipage}\hspace{0cm}
	\begin{minipage}[b]{0.49\linewidth}
		\includegraphics[width=\linewidth]{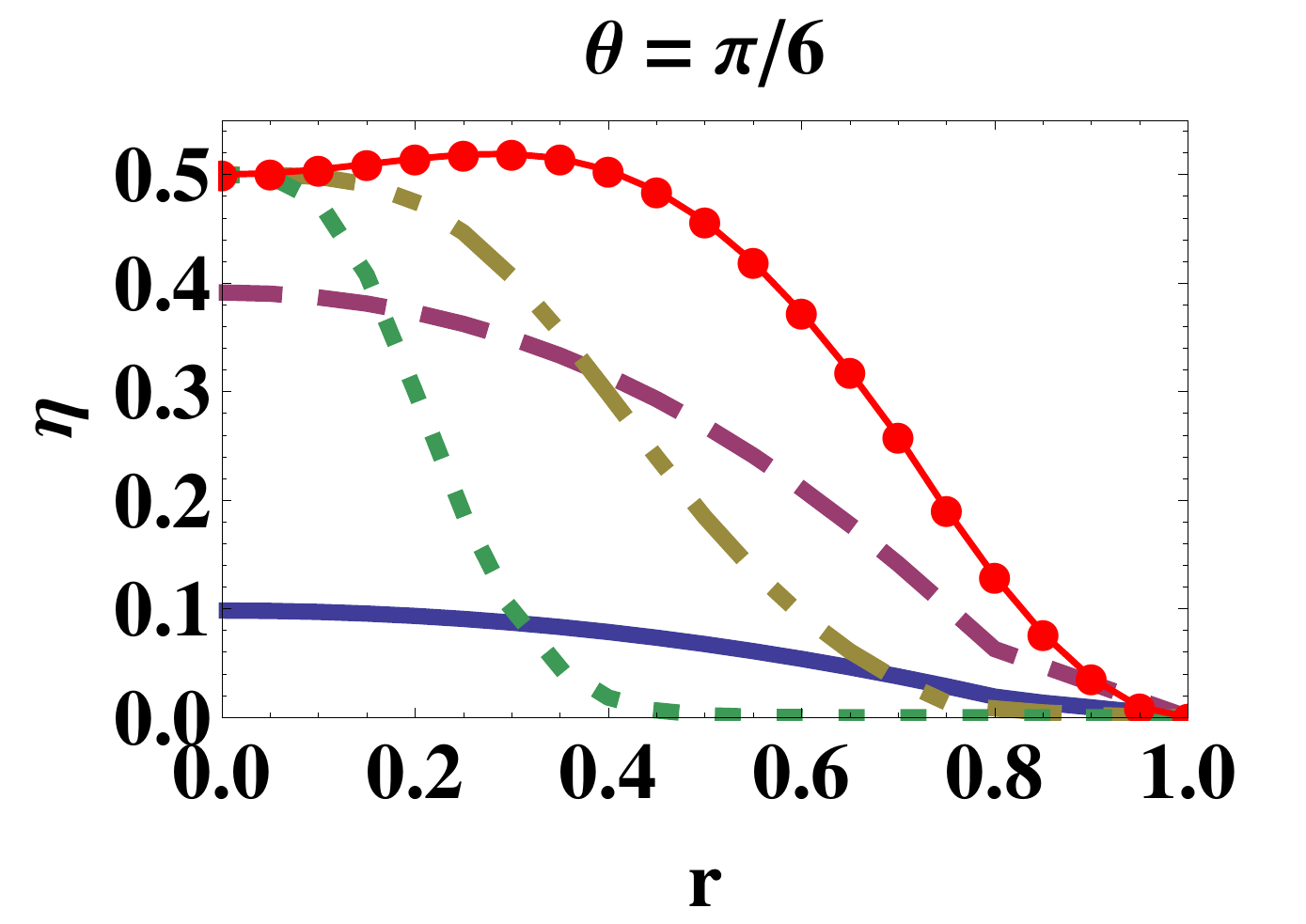}
	\end{minipage} 
	\begin{minipage}[b]{0.49\linewidth}
		\includegraphics[width=\linewidth]{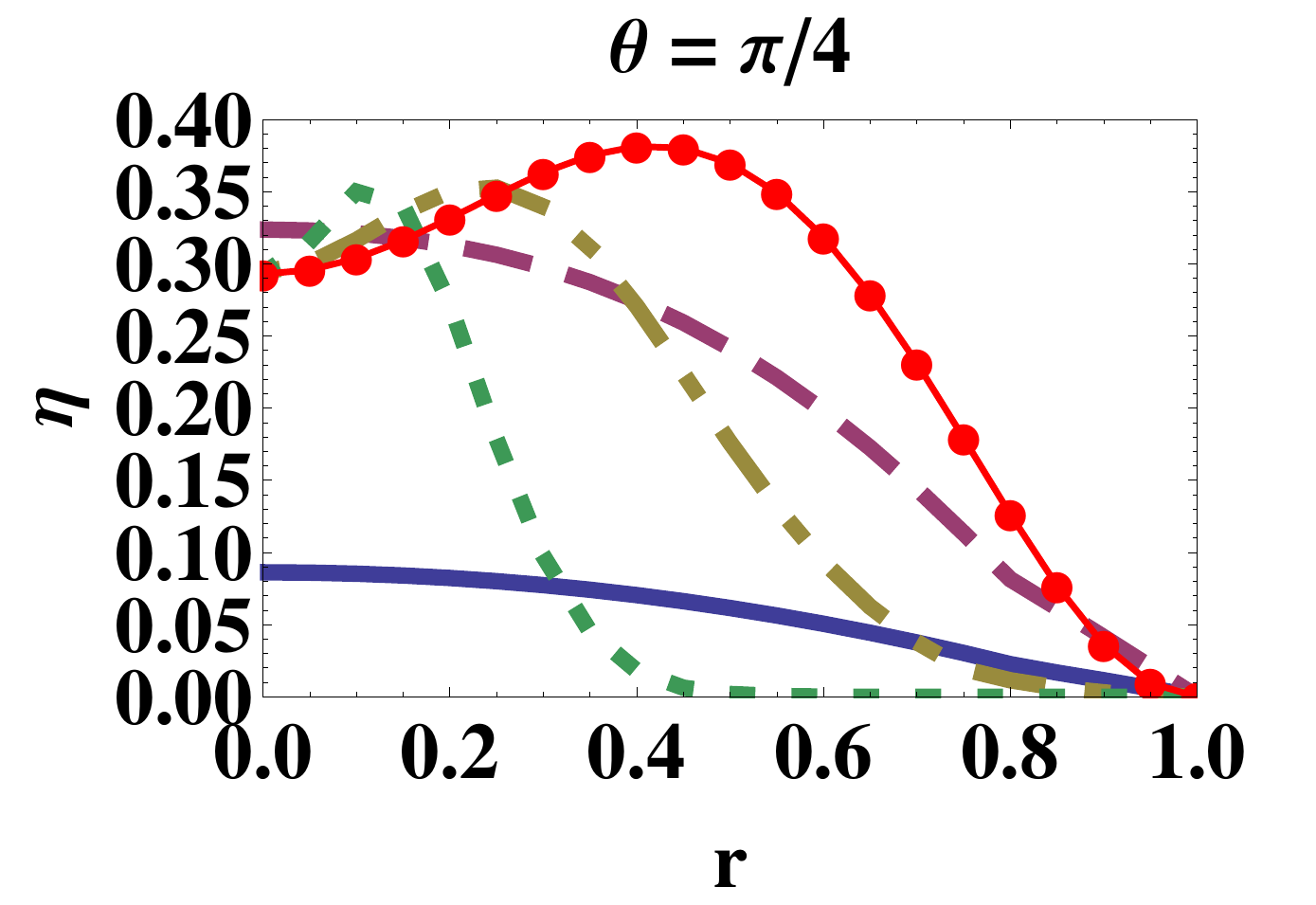}
	\end{minipage}\hspace{0cm}
	\begin{minipage}[b]{0.49\linewidth}
		\includegraphics[width=\linewidth]{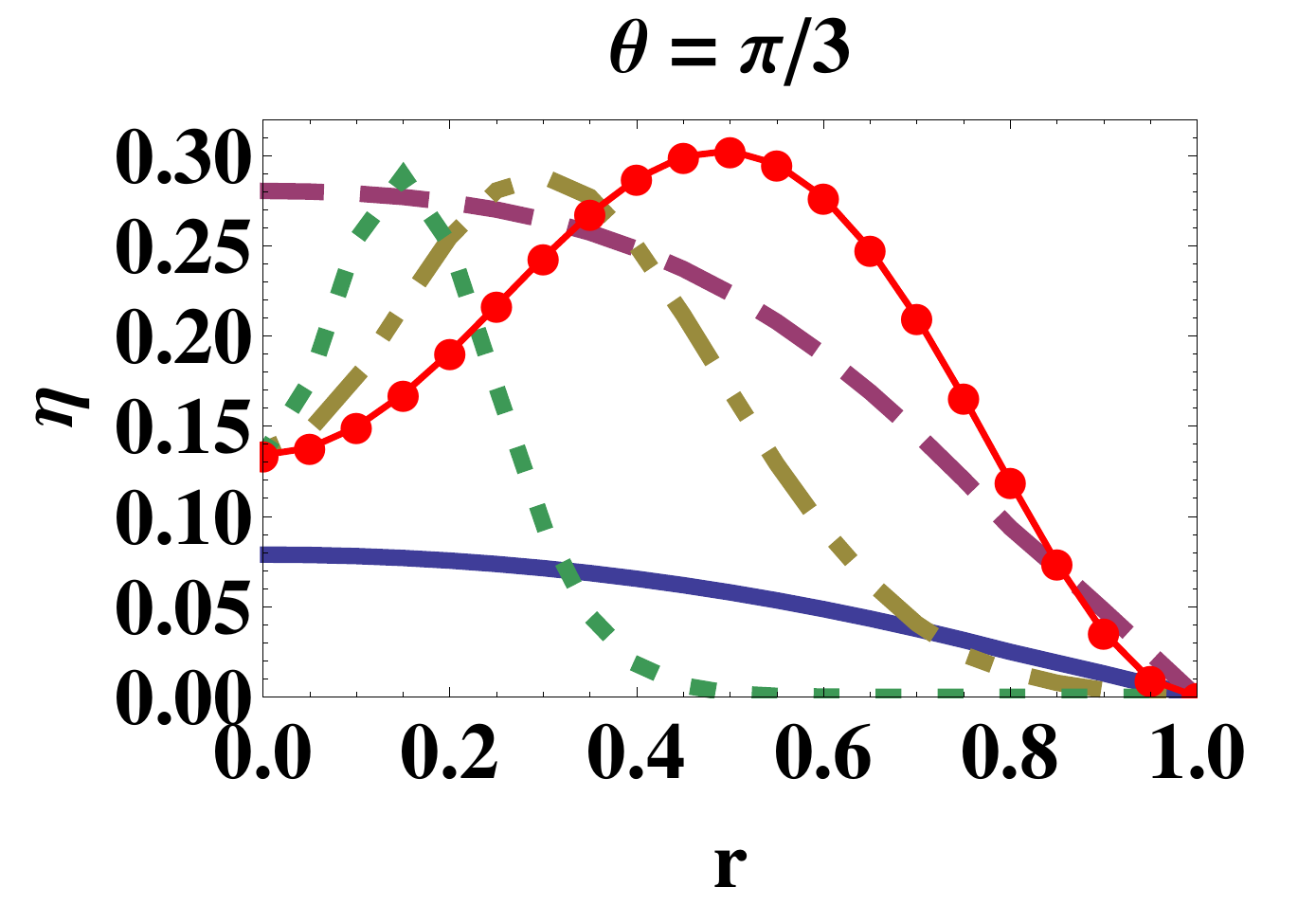}
	\end{minipage}
	\vspace{-0.2cm}
	\caption{(Color online) Efficiency of the CT protocol as a function of normalized time for the VSP and coherent states encoding and different values of $\theta$. Line with circles is used to the VSP case. For the coherent states, we have $\alpha=0.20$ (solid), $\alpha=0.50$ (dashed), $\alpha=1.25$ (dot-dashed), $\alpha=2.50$ (dotted).}
	\label{efi}
\end{figure}

In Fig.\ref{efi}, we present the plots of efficiency for both physical encodings: VSP in Eq.(\ref{k1}) and coherent state in Eq.(\ref{k2}). Our goal now is to compare the performance of these two kind of photonic qubits in the CT protocol under the scenario depicted in Fig.\ref{esquema}. According to Fig.\ref{efi}, it is the VSP encoding with $\theta=0$ (GHZ state) that presents the best efficiency. As $\theta$ is varied, however, crossings may happen involving the VSP and the coherent states encoding. For instance, for $\theta=\pi/3$  it is quite clear that for weak or moderate losses $r\leq 0.3$ the coherent states do a better job than the VSP states. Once again, the use of small amplitude or large amplitude coherent states will depend on the precise value of $r$ one has, i.e., the quality/length of the optical fibers. Finally, it worthwhile to notice that the efficiency is not, in general, a monotonically decreasing function of the normalized time what might be very useful for practical purposes. This happens because the efficiency defined in Eq.(\ref{eficiencia}) takes into account, on an equal footing,  the twofold aspects of the CT protocol: control power and conditioned fidelity.
\subsection{Tripartite non-locality}
The relation between non-locality, in the sense of violation of Bell inequalities, and teleportation has a long history in quantum information \cite{popescu}. In the bipartite scenario of standard teleportation, it can be shown that violation of a Bell inequality by the state shared by Alice and Bob implies that the state is useful for quantum teleportation (provides fidelities above the classical limit $2/3$) \cite{horodecks2}. However, no violation does not imply that the state is useless for quantum teleportation \cite{popescu}. When looking at more general scenarios, it can be shown that any bipartite state which is useful for teleportation is also a nonlocal resource \cite{multcopy}. This means that a Bell inequality is always violated deterministically when a sufficiently large number of copies of the state is provided. For the tripartite scenario of the CT protocol, there is yet no general results linking legitime tripartite non-locality and the performance of the protocol. In order to shed some light on this problem, we will now study violation of tripartite non-locality and its possible (or lack of) relation with the efficiency of the CT protocol. 

 One possibility is the use of the Bell-Svetlichny inequality which was the first tool proposed to test genuine multipartite non-locality \cite{Svetlichny}. Suppose a set o dichotomic observables $A_2$ ($A'_2$), $B_3$ ($B'_3$), $C_1$ ($C'_1$), where $A_2$ is an observable acting on the state space of qubit $2$ (Alice's laboratory), $C'_1$ is an observable acting on the state space of qubit $1$ (Charlie laboratory), etc. At each run of the experiment, each party chooses one of the two observables at random to measure. For dichotomic observables with spectrum $\pm1$, the Bell-Svetlichny inequality can be written as \cite{cereceda}
\begin{align}
	|S_v|=|&E(A,B,C)+E(A,B,C')+E(A,B',C)\nonumber\\
	+&E(A',B,C)-E(A',B',C')-E(A',B',C)\nonumber\\
	-&E(A',B,C')-E(A,B',C')|\leq 4,
	\label{svdes}
\end{align}
where $E(A,B,C)\equiv {\rm{Tr}}\{\rho O\}$ is the expectation value of the three-body operator $O=ABC$ and $\rho$ is the tripartite state whose nonlocal features one is interested in. We call $S_v$ the Bell-Svetlichny function. The GHZ state is known to maximally violate this inequality with $|S_v|=4\sqrt{2}$ \cite{Svetlichny,tripart1}. It is important to have in mind that, if one state does not violate a given inequality, it does not immediately imply that it will not violate other inequality built with a different set of dichotomic observables \cite{rmp}.

For the VSP-encoded MS state evolving according to Eq.(\ref{BMeq}), the three-body operator $O$ we will be using to evaluate $S_v$ is
\begin{align}
O_{\rm{VSP}}=\Omega_1(\chi)\Omega_2(\lambda)\Omega_3(\mu),
\label{OII}
\end{align}
where each dichotomic operator is defined as
$\Omega_j(\zeta)=R_j(\zeta)\sigma_z{}_jR_j{}^\dagger(\zeta)$, with $\sigma_z{}_j$ the usual Pauli-$z$ operator acting on the state space of qubit $j$ ($j=1,2,3$), and $R_j(\zeta)$ a rotation operator acting on the same space. In the Pauli-$z$ basis it reads
\begin{align}
&R_j(\zeta)=
\begin{pmatrix}
\cos|\zeta| & \frac{\zeta}{|\zeta|}\sin|\zeta|\\
-\frac{\zeta^*}{|\zeta|}\sin|\zeta| & \cos|\zeta|
\end{pmatrix},
\end{align}
where $\zeta=-\omega e^{-i\delta}/2$ with $0\leq\omega\leq\pi$ and $0\leq\delta\leq2\pi$. It is worth noticing that by choosing $\zeta$ complex, our maximization is more apt to detect stronger violations \cite{pj}. Finally, in the notation used in Eq.(\ref{svdes}), one has $A_2=\Omega_2(\lambda)$, $A'_2=\Omega_2(\lambda')$, $B_3=\Omega_3(\mu)$, and so forth.

\begin{figure}[!htb]
	\begin{minipage}[b]{0.49\linewidth}
		\includegraphics[width=\linewidth]{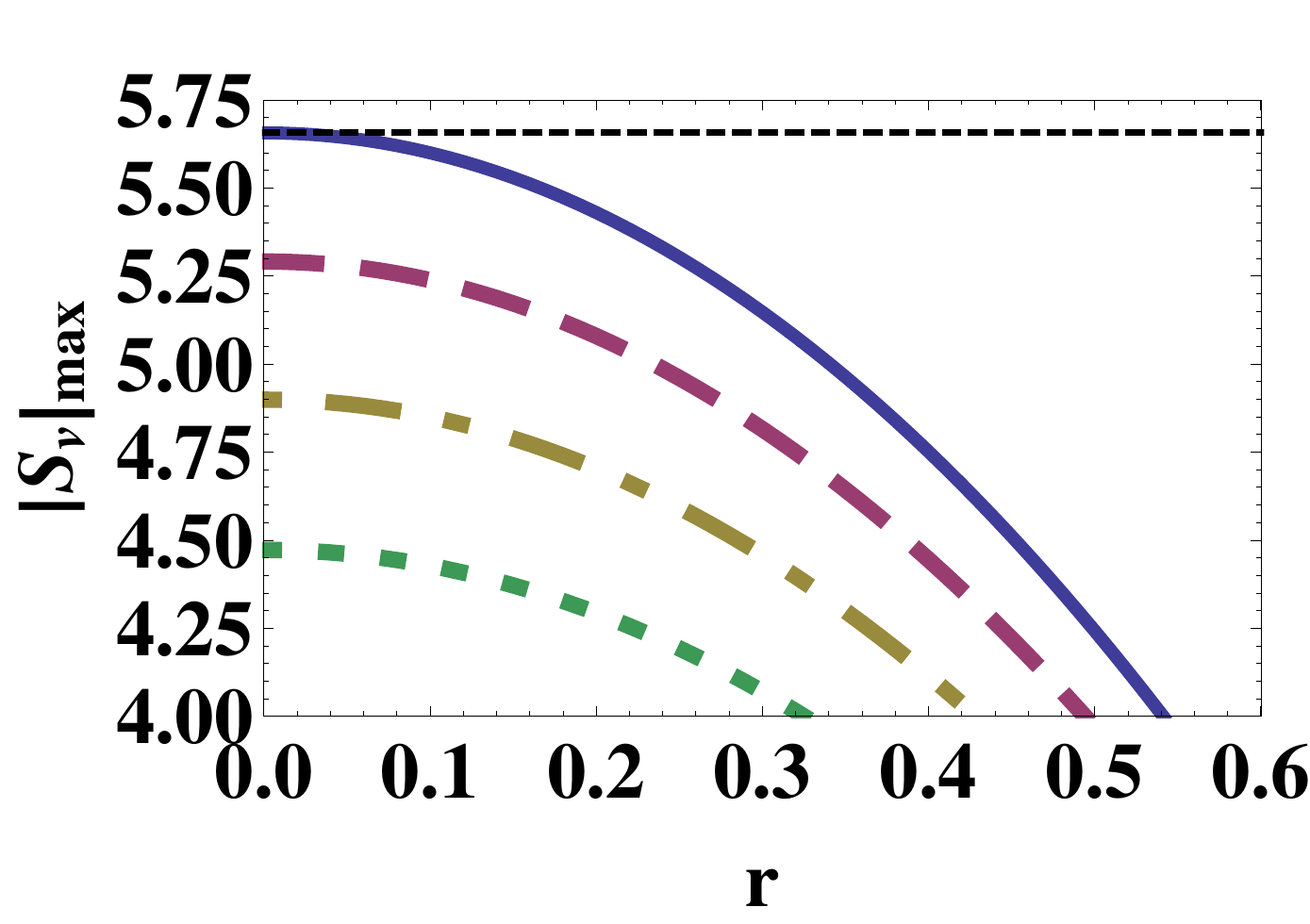}
	\end{minipage}\hspace{0cm}
	\begin{minipage}[b]{0.49\linewidth}
		\includegraphics[width=\linewidth]{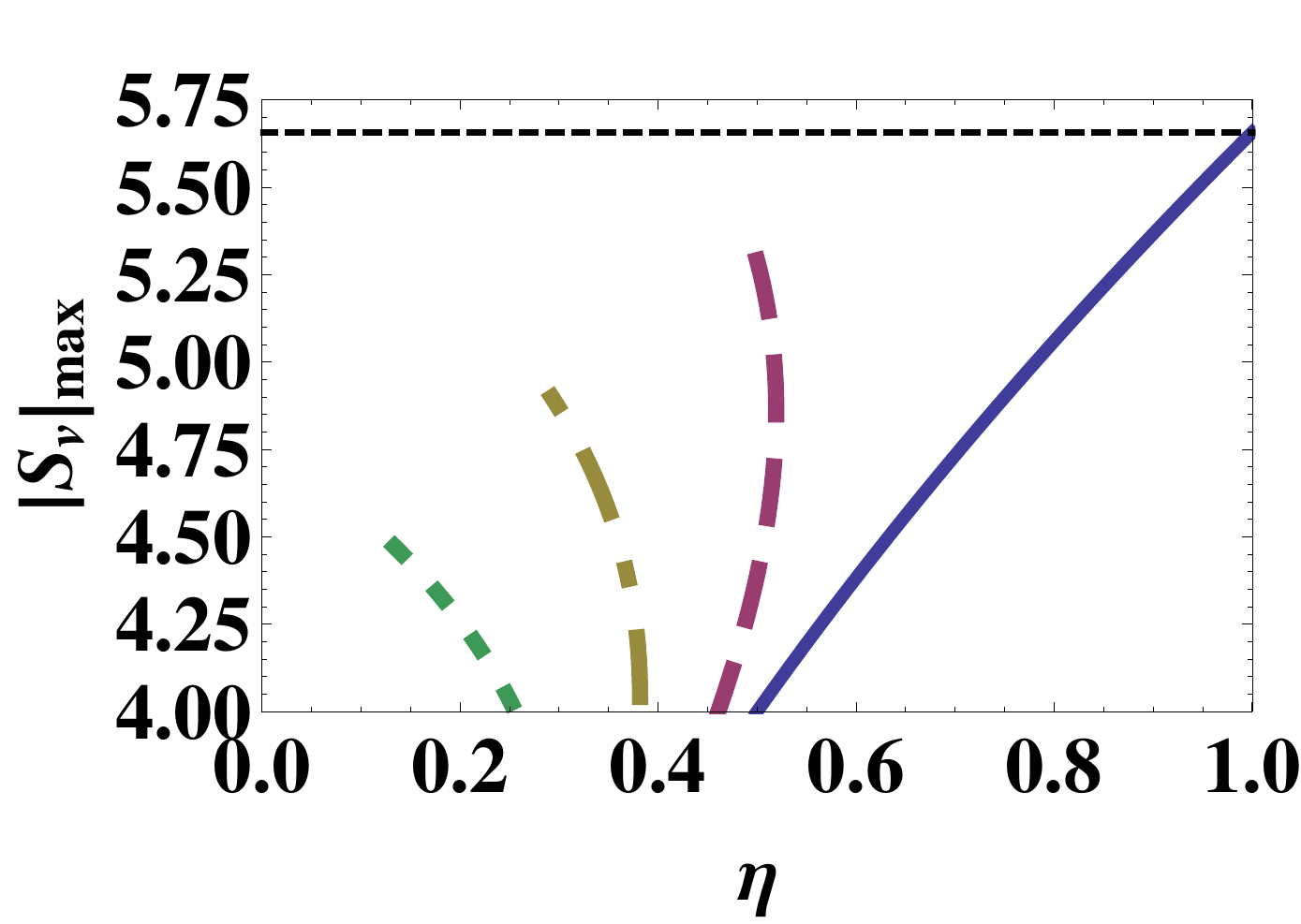}
	\end{minipage} 
	\vspace{-0.2cm}
	\caption{(Color online) Left: Maximized Bell-Svetlichny function $|S_{v}|_{\rm{max}}$ for the VSP encoding Eq.(\ref{k1}) as a function of the normalized time and for different values of $\theta$. Right: Maximized Bell-Svetlichny $|S_{v}|_{\rm{max}}$ against efficiency $\eta$ as the normalized time varies in the interval $[0,1]$. Only the region where the Bell-Svetlichny inequality is violated that is shown. We used $\theta=0$ (solid), $\theta=\pi/6$ (dashed), $\theta=\pi/4$ (dot-dashed) e $\theta=\pi/3$ (dotted). The dotted straight line indicate maximal violation of the Bell-Svetlichny inequality allowed by quantum mechanics.} 
	\label{bellvsp}
\end{figure}

In Fig.\ref{bellvsp}, we present $|S_v|$ maximized over the twelve parameters that comprise the two measurements each party can choose. In the left panel, we see how the maximal Bell-Svetlichny function evolves under dissipation. For the ideal fiber $r=0$, the maximally entangled GHZ ($\theta=0$) indeed maximally violates the Bell-Svetlichny inequality. Also at $r=0$, as $\theta$ increases in the range $[0,\pi/2]$, the violation becomes progressively smaller. This can be understood by looking at the tripartite entanglement of the pure MS state in Eq.(\ref{MS1}) as quantified by the tangle $\Upsilon$ \cite{tangle}. For this state, $\Upsilon=\cos^2\theta$ \cite{msstate}, i.e., it is a monotonically decreasing function of $\theta$ as it varies from $0$ to $\pi/2$. The observed decay with $r$ is the expected effect of the losses which, in general, depletes the quantum features.

In the right panel of Fig.\ref{bellvsp}, we present parametric plots where each point corresponds to $(\eta(r),|S_{v}|_{\rm{max}})$. Again, only the nonlocal region is displayed. From this plot, it is possible to see that $|S_{v}|_{\rm{max}}$ and $\eta$ are not trivially related. For the GHZ ($\theta=0$), the better the efficiency the stronger the non-locality and vice versa. However, for finite  $\theta$, there can be situations where the efficiency growth is accompanied by the $|S_{v}|_{\rm{max}}$ depletion and the other way round.   

For the coherent states, it is necessary to adapt the three-body operator $O$ to an hybrid scenario \cite{pj}. We need now dichotomic operators acting on the whole state space of each field mode and not just on a subspace of two states as in the VSP. In this work, we will employ
\begin{align}
O_{\rm{coh}}=\Omega_1(\chi)\Pi_2(\lambda)\Pi_3(\mu),
\label{OI}
\end{align}
where  $\Pi_j(\beta)$ is the displaced parity operator \cite{parity}
\begin{align}
\Pi_j(\beta)=D_j(\beta)\sum_{n=0}^{\infty}(|2n\rangle_j\langle 2n|-|2n+1\rangle_j\langle 2n+1|)D_j^\dagger(\beta),
\end{align}
with $D_j(\beta)$ the displacement or Glauber operator \cite{iqopt} acting on modes $j$ ($j=2,3$), $\beta$ a complex number, and $\ket{n}_j$ Fock states of mode $j$. For qubit $1$, possessed by Charlie, we keep using the same dichotomic operator as in the VSP case giving that he still performs the projection on the discrete set in Eq.(\ref{base}).

\begin{figure}[!htb]
	\begin{minipage}[b]{0.49\linewidth}
		\includegraphics[width=\linewidth]{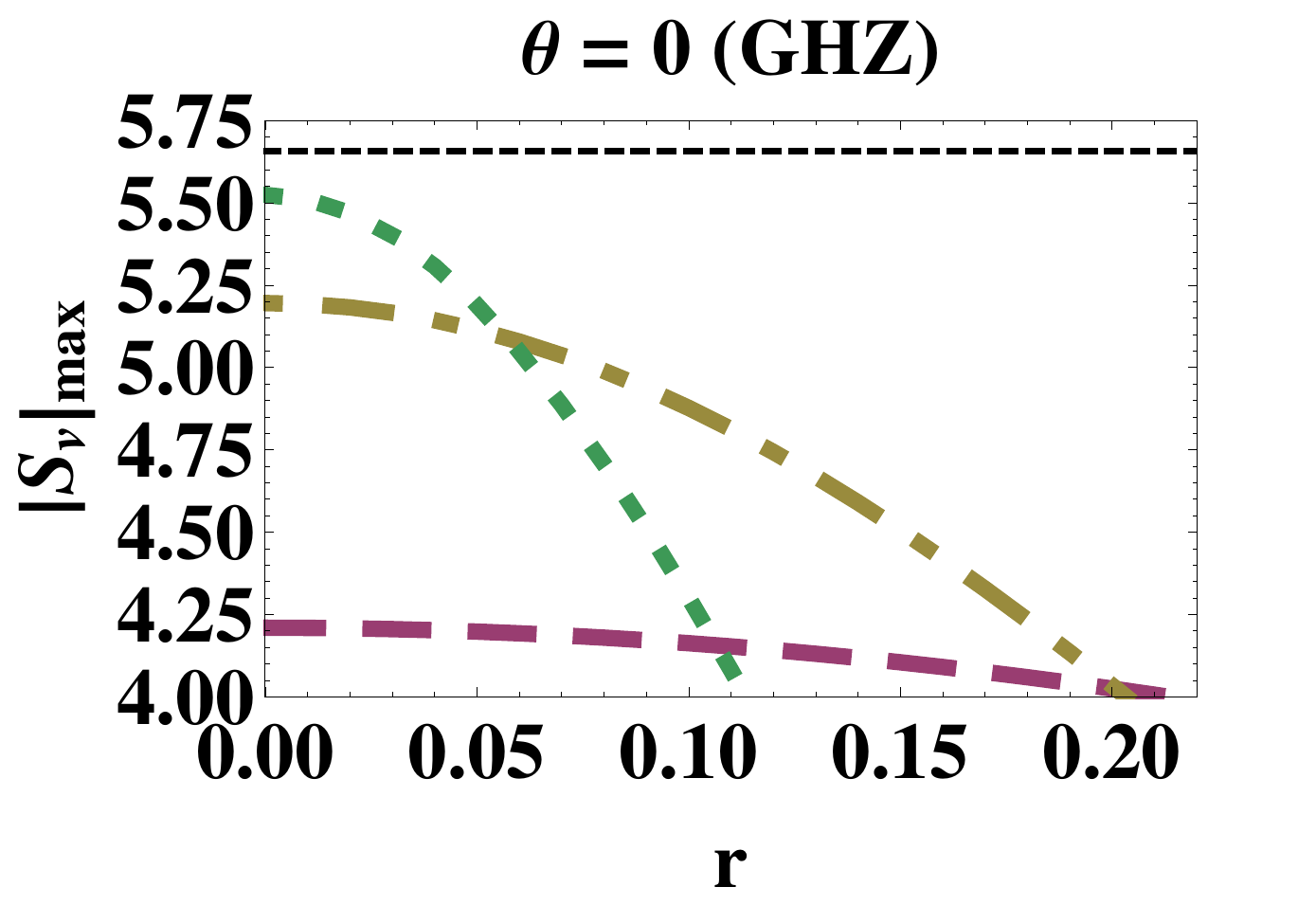}
	\end{minipage}\hspace{0cm}	
	\begin{minipage}[b]{0.49\linewidth}
		\includegraphics[width=\linewidth]{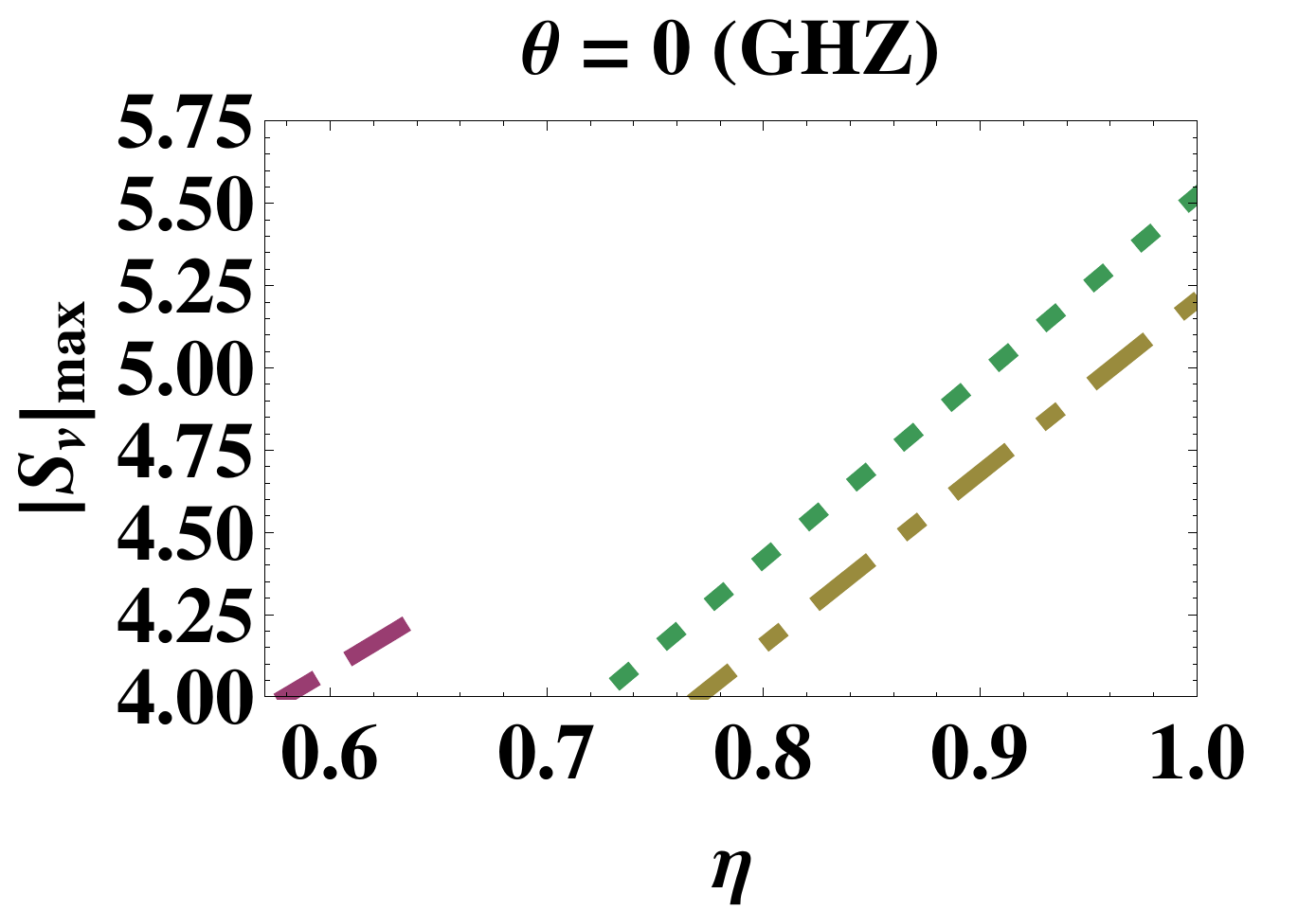}
	\end{minipage}\hspace{0cm}
	\begin{minipage}[b]{0.49\linewidth}
		\includegraphics[width=\linewidth]{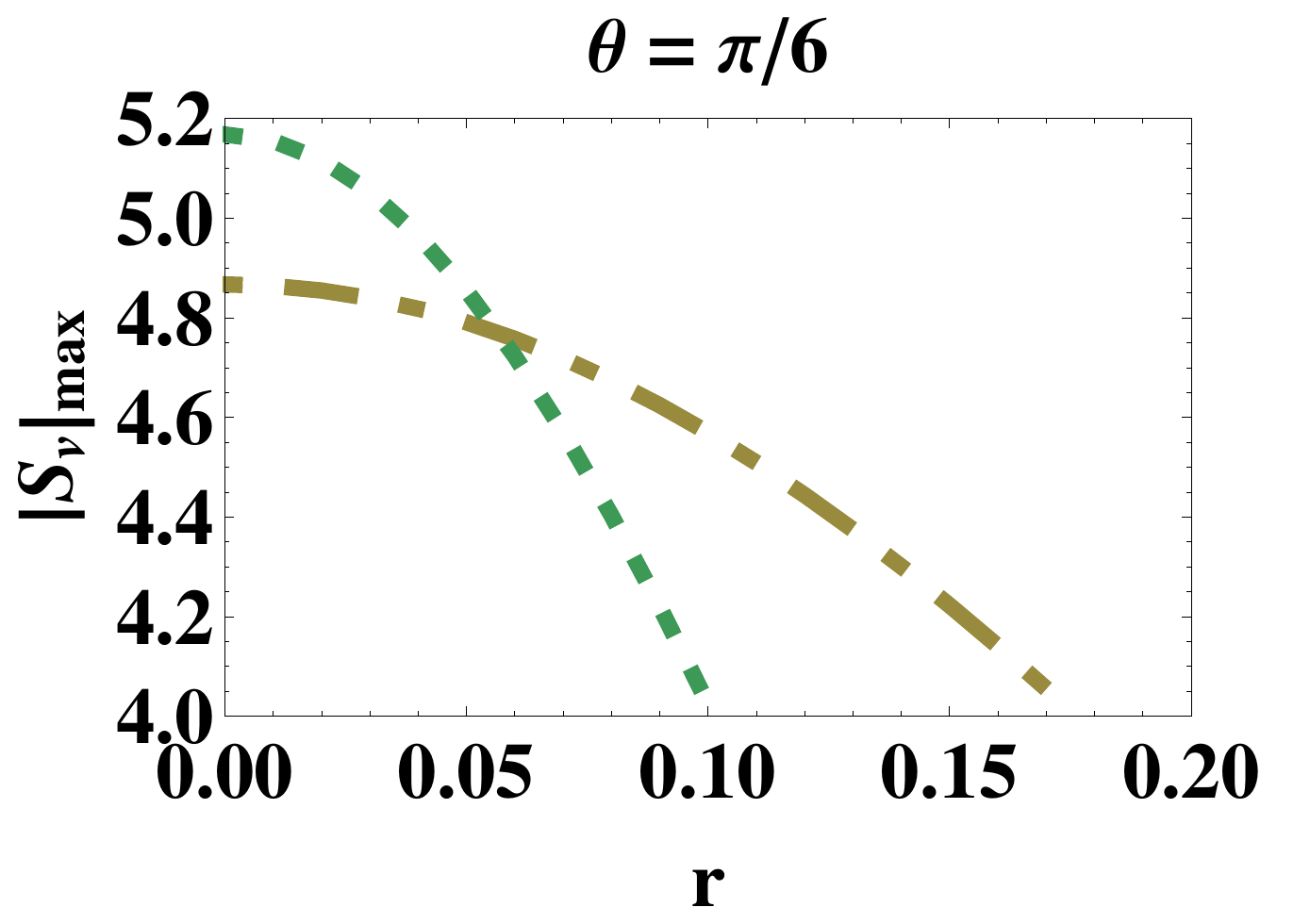}
	\end{minipage} 		
	\begin{minipage}[b]{0.49\linewidth}
		\includegraphics[width=\linewidth]{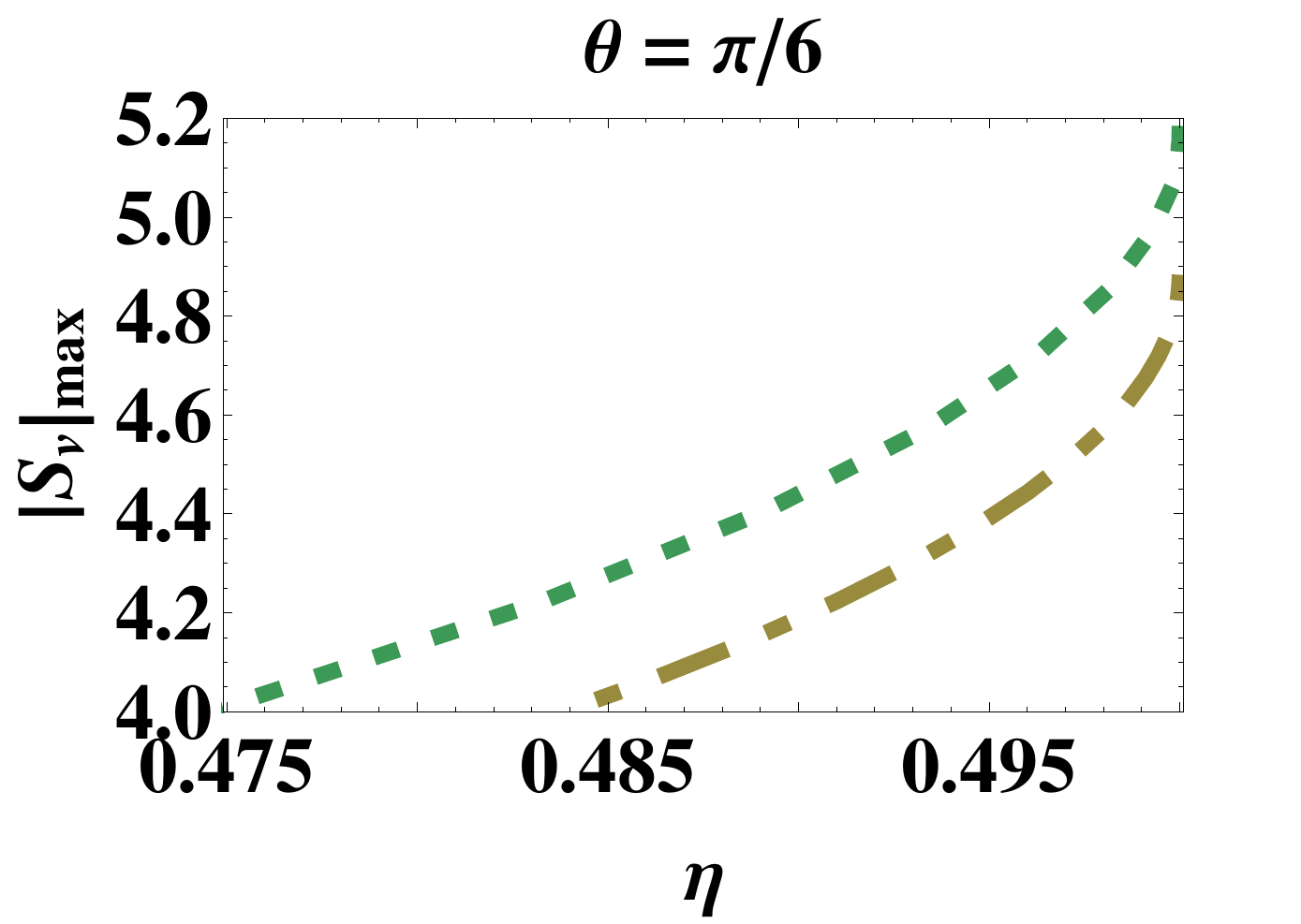}
	\end{minipage} 
	\begin{minipage}[b]{0.49\linewidth}
		\includegraphics[width=\linewidth]{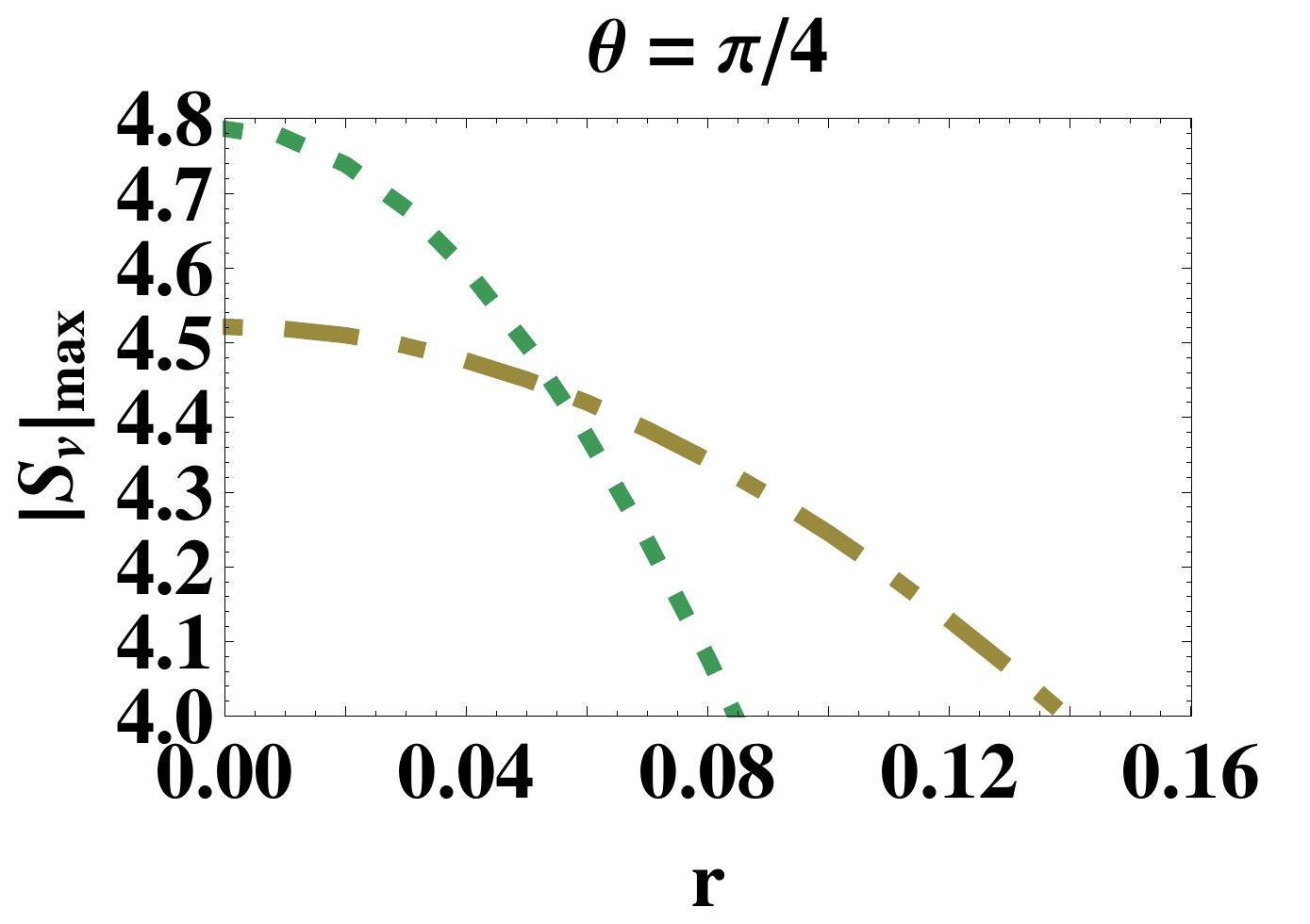}
	\end{minipage}\hspace{0cm}	
	\begin{minipage}[b]{0.49\linewidth}
		\includegraphics[width=\linewidth]{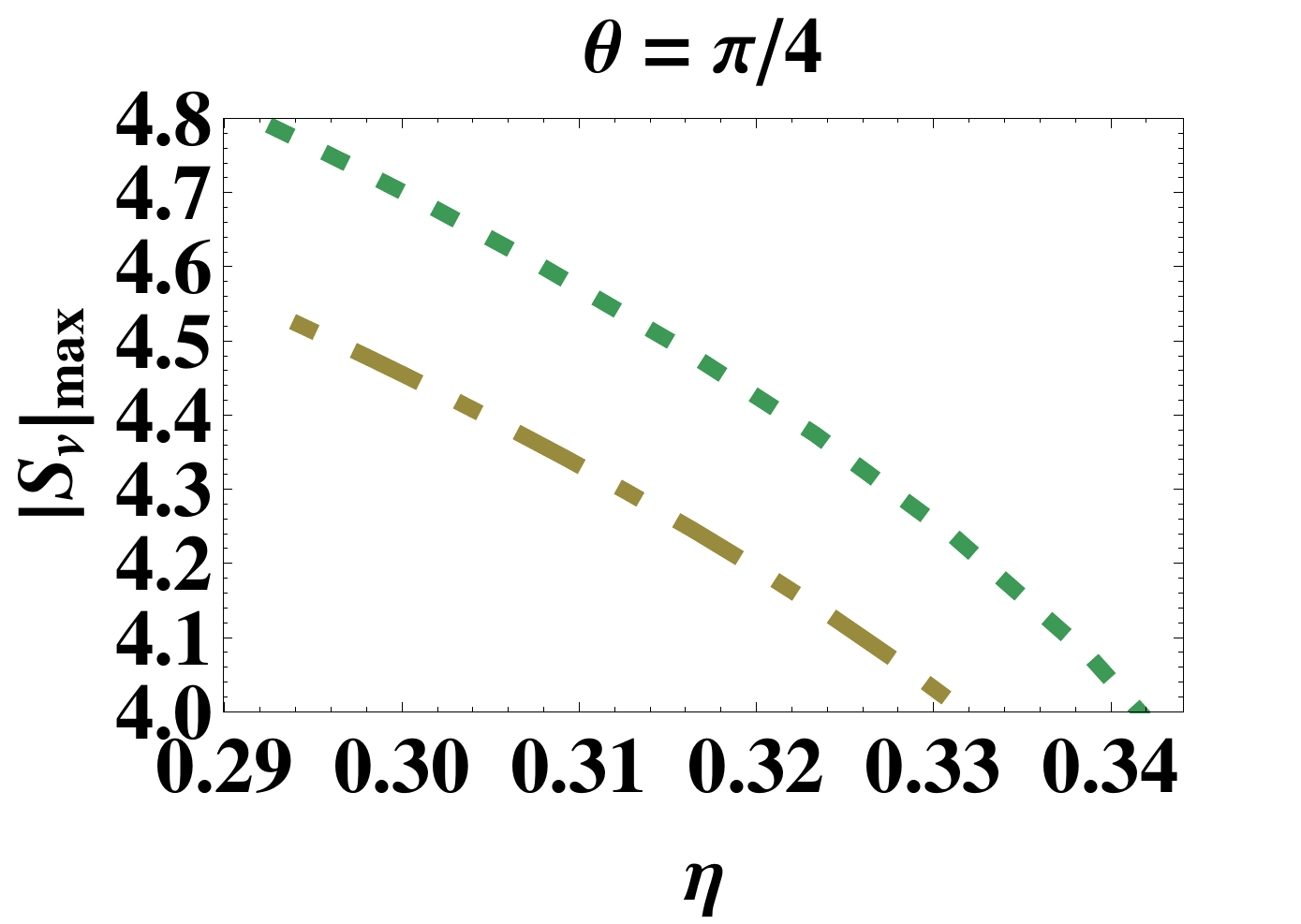}
	\end{minipage}\hspace{0cm}
	\begin{minipage}[b]{0.49\linewidth}
		\includegraphics[width=\linewidth]{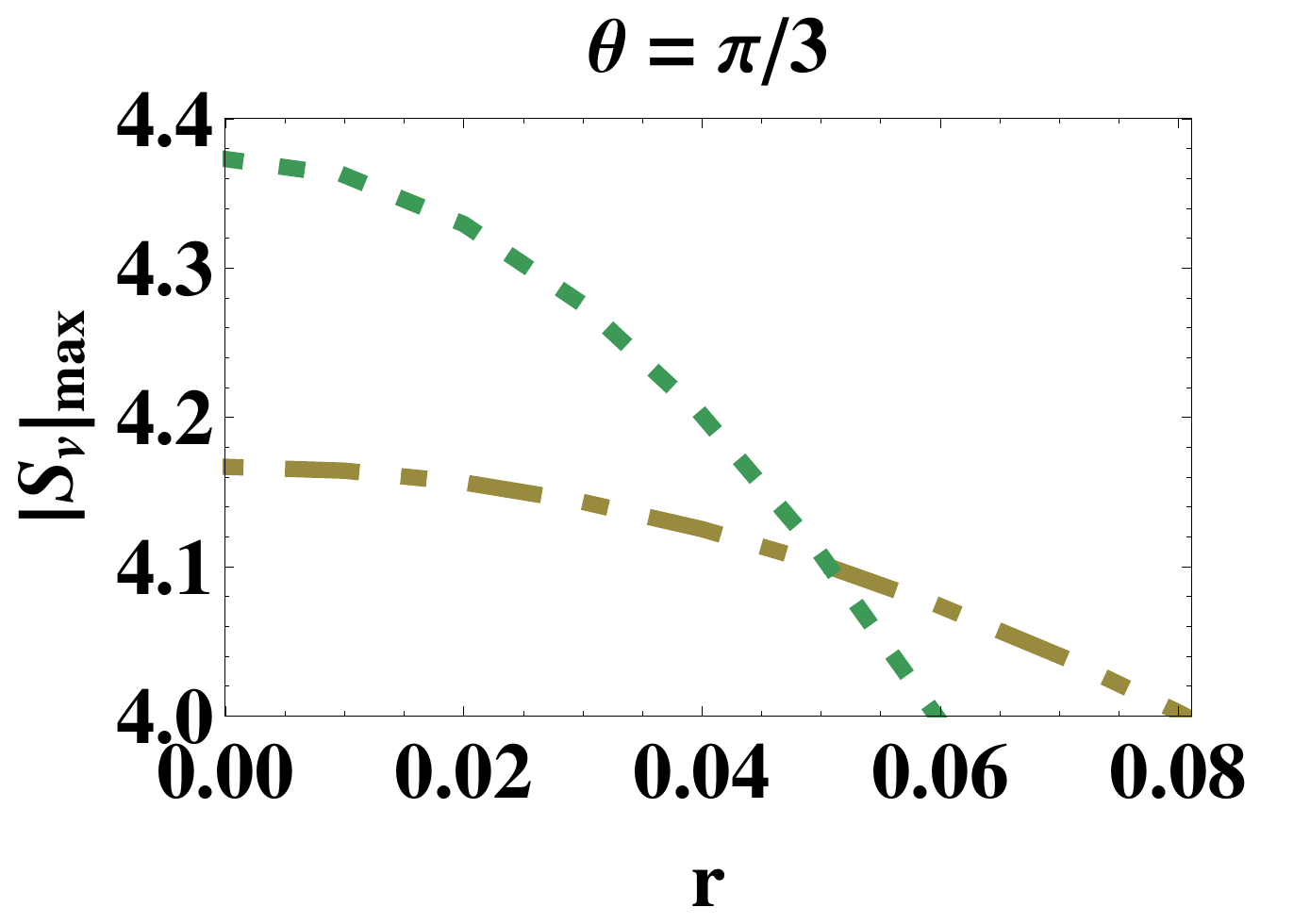}
	\end{minipage}	
	\begin{minipage}[b]{0.49\linewidth}
		\includegraphics[width=\linewidth]{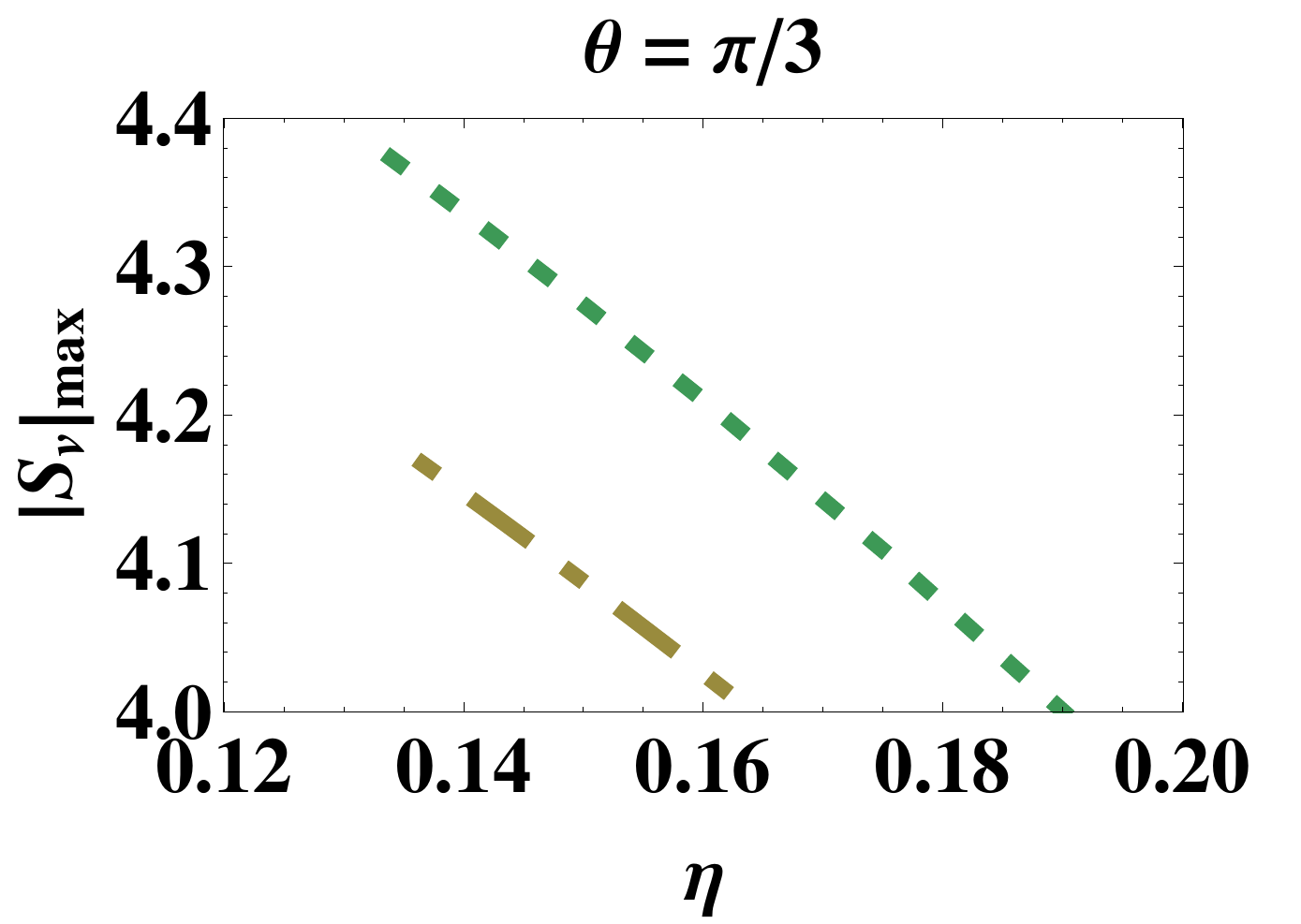}
	\end{minipage}
	\vspace{-0.2cm}
	\caption{(Color online) Left: Maximized Bell-Svetlichny function $|S_{v}|_{\rm{max}}$ for the coherent state encoding Eq.(\ref{k2}) as a function of the normalized time and for different amplitudes $\alpha$. Right: Maximized Bell-Svetlichny $|S_{v}|_{\rm{max}}$ against efficiency $\eta$ as the normalized time varies in the interval $[0,1]$. Only the region where the Bell-Svetlichny inequality is violated that is shown. We used $\alpha=0.50$ (dashed), $\alpha=1.25$ (dot-dashed), $\alpha=2.50$ (dotted). The dotted straight line on the top panels indicate maximal violation of the Bell-Svetlichny inequality allowed by quantum mechanics.}
	\label{bellcs}
\end{figure}

Again, the Bell-Svetlichny function will depend on twelve variables because the argument of each operator in Eq.(\ref{OI}) is a complex number. In Fig.\ref{bellcs}, we present its maximization as a function of normalized time (left panels) and the parametric plots $(\eta(r),|S_{v}|_{\rm{max}})$ (right panels). We vary the amplitude $\alpha$ of the coherent states in Eq.(\ref{k2}) as well as the parameter $\theta$ which defines the MS state in Eq.(\ref{MS1}).
One can see that, except for $\theta=0$, only the states with large amplitude violate of the Bell-Svetlichny inequality. Also, there are cases where weak violations provide better efficiencies than strong violations [see, for instance, the case $\theta=\pi/4$]. The Bell-Svetlichny function, like the efficiency, control power and conditioned fidelities, also presents crossings. Finally, from Figs. (\ref{efi}), (\ref{bellvsp}), and (\ref{bellcs}), one can see that the non-locality is more affected by losses than the performance of CT protocol. In other words, even after $|S_v|$ starts assuming values compatible with a classical local and realist theory, the CT protocol still produces results compatible only with quantum theory. In this sense, for the physical setting described in Fig.\ref{esquema}, the CT protocol is more robust against dissipation than the non-locality quantified using Bell-Svetlichny.

\section{Final remarks}\label{conc}
In this work, we presented a careful study of the controlled teleportation protocol with two of the most used optical qubits in the single-rail logic: superpositions of vacuum and single-photon  (VSP) states and superposition of coherent states of opposite phases. The analysis took into account losses in the fibers used by the controller to distribute the optical qubits to the other parties. We introduced an efficiency quantifier which takes into account the main features of the protocol. We found that the best performance is achieved with the VSP encoding and the GHZ state, which is a proper limit of the MS states considered in this work. For general MS states, however, the losses affect the efficiencies for the VSP and the coherent states in a non trivial way. Depending on strength of the losses, either the VSP or the coherent states might be the best choice. We also investigated a possible relation between tripartite non-locality, under the scope of the Bell-Svetlichny inequality, and the efficiency of the protocol. We found that there is no simple or monotonic relation between these physical quantities. 

It is important to remark that the teleportation fidelities in this work are evaluated assuming that Bob is able to perform a set of unitary rotations on his qubit, as demanded by the standard quantum teleportation protocol \cite{tp}. The optical elements demanded to perform such rotations are well known since the very first proposals for the use of coherent states in quantum information tasks \cite{jeong1,jeong2,milburn}.  On Alice side, the Bell measurement she performs is typically  probabilistic when using coherent states. This is a consequence of the non-orthogonality of the basis states. Bell measurements with coherent states are carefully discussed in  \cite{hir,gege}. Simple schemes may give success probabilities approaching $50\%$ as discussed in \cite{hir}. The probabilistic nature of the Bell measurements does not affect our calculations giving that only the successful events have been considered. Details in the appendix.   

Finally, it would be interesting to expand our analysis of losses and optical encodings to the multiqubit scenario discussed in \cite{mq1,mq2}. In particular, it may be the case that the interplay between dissipation and non-orthogonality, appearing in the coherent states,  may have consequences in the minimal control power \cite{mq2}. Additionally, as damping impacts the knowledge the controller has on the channel (changes the state entropy), it might be of interest to include losses in the discussion proposed in \cite{cr}.

We hope our work may be useful to a better understanding of the controlled teleportation protocol and motivate its future experimental implementation with optical qubits. 

\acknowledgments 
IM acknowledges support by the Coordenação de Aperfeiçoamento de Pessoal de Nível Superior (CAPES). FLS acknowledges partial support from CNPq (grant nr. 307774/2014-7).
\section*{ Appendix } \appendix                                       
\subsection{Teleportation Fidelity} \label{app}    
Let us suppose that Alice and Bob share a general bipartite state described by the density operator $\rho$, and they want to use this state as a channel for teleportation \cite{tp}. Assuming that Alice \textit{succeeds} in projecting the local state of her two qubits in a Bell state, and that Bob can perform arbitrary rotations on his own qubit,  it can be shown that the maximal fidelity for the standard teleportation protocol can be written as  \cite{horodecks1}
\begin{align}
F(\rho)=\frac{2 f + 1}{3},
\end{align}
with $f$ being the so-called fully entangled fraction defined as \cite{fmax}
\begin{align}
f=\underset{\ket{\phi}}{{\rm{max}}}\bra{\phi}\rho\ket{\phi},
\end{align}
where the maximization is over all maximally entangled states $|\phi\rangle$, i.e., all states that can be obtained from a singlet using local unitary transformations. 

The fully entangled fraction can be found using a simple method whose steps we now briefly describe \cite{james}. First , the density operator $\rho$ must be written in a special basis usually referred to as the \textit{magic basis}
\begin{align}
&\ket{m_1}=\ket{\Phi^+}=\frac{1}{\sqrt{2}}(\ket{00}+\ket{11}),\nonumber\\
&\ket{m_2}=i\ket{\Phi^-}=\frac{i}{\sqrt{2}}(\ket{00}-\ket{11}),\nonumber\\
&\ket{m_3}=i\ket{\Psi^+}=\frac{i}{\sqrt{2}}(\ket{01}+\ket{10}),\nonumber\\
&\ket{m_4}=\ket{\Psi^-}=\frac{1}{\sqrt{2}}(\ket{01}-\ket{10}). \nonumber
\end{align} 
The fully entangled fraction $f$ is then evaluated simply as the biggest eigenvalue of the real part of  $\rho$ when written in the magic basis \cite{james}.
%

\end{document}